\documentclass[10pt]{article} 

\usepackage{amsfonts,amsmath,amssymb}
\usepackage{graphicx,verbatim,hyperref}
\usepackage[frak=esstix,scr=boondoxo,cal=dutchcal]{mathalfa}
\usepackage{natbib}
\def\citep{}\renewcommand{\citep}[2][]{(\citealt[#1]{#2})}

\def\sun{\odot} 
\def\sfrac#1#2{\genfrac{}{}{}{1}{#1}{#2}}
\def\top#1#2{\genfrac{}{}{0pt}{3}{\raisebox{-1.9pt}{\(\scriptscriptstyle #1\)}}{\raisebox{1.9pt}{\(\scriptscriptstyle #2\)}}}

\begin{document}
\begin{center} 
{\Large\bf The Astrophysical Corona as the Minimum Atmosphere Surrounding
Embedded Non-Force-Free Flux Tubes}
\vskip5mm
{\bf Laurence J. November}\\
\vskip2mm
La Luz Physics, La Luz NM 88337-0217 USA\\
{\it laluzphys@yahoo.com}\\
March 26, 2019
\end{center}

\begin{abstract}
\noindent 
The equilibrium of current-carrying magnetic fields (e.g.\ flux tubes) embedded
in a large-scale background field is developed and discussed in the
astrophysical context.  Embedded non-force-free current-carrying fields require
a minimum surrounding atmosphere, which by direct pressure balance has a gas
pressure everywhere proportional to the background magnetic pressure.  Formally,
the MHD equations, with flows and gravity as part of a wide class of physical
processes, separate into independent local and global relations representing an
equilibrium solution for embedded current-carrying fields. The local pressure
relation for the embedded field is a 3D Grad-Shafranov equation with
finite-sheath solutions.  The global relation reproduces the ambient MHD
pressure equation without the embedded fields, but instead with the constraint
that the ambient gas and magnetic pressures vary in proportion, as with the
direct pressure balance.  A coupled gas pressure in magnetically dominant
regimes necessitates refilling outflows in a depleted atmosphere (actualized by
flux-tube Lorentz forces) providing a compressively heated equilibrium corona
with a specific global distribution of density, temperature, and steady
accelerated outflow, all defined by the large-scale background magnetic field.
Magnetic footpoint compression and twisting in a high-gas-pressure field-forming
region (e.g.\ convection zone) outside, as below, the magnetically dominant
regime, can introduce and sustain non-force-free embedded fields, thereby
providing the energy for the coronal atmosphere. Such coronae may be relevant on
very different astrophysical scales:\ around the sun and stars, and ranging from
planets, to neutron stars, black holes, and spiral galaxies. Predicted coronal
temperatures are corroborated.
\vskip2mm\noindent{\bf keywords:} magnetohydrodynamics (MHD) -- Sun: corona --
Sun:  transition region -- solar wind -- stars:  coronae -- planets and
satellites: atmospheres -- galaxies: fundamental parameters
\end{abstract}

\section{Introduction}
\label{s:intro}

A spatially coherent magnetic field strewn with isolated fine filaments appears
to be a usual form occurring on a number of very different astrophysical scales.
The solar corona, a well-observed example, contains much fine filamentary
structure, thought to be evidence for magnetic flux tubes embedded in its strong
background magnetic field.  Fine flux tubes exhibit a constant pressure contrast
all along their lengths through the solar corona from evacuated to twice the
ambient, consistent with expected internal magnetic displacement over- and
under-pressures, respectively, in non-force-free magnetic fields. The
observational evidence for non-force-free flux tubes in the solar corona is
reviewed in Section \ref{s:obs}.

The apparent existence of non-force-free flux tubes embedded in the large-scale
background solar magnetic field leads to the didactic question:  If the external
ambient gas pressure were to decrease somewhere along the length of an evacuated
non-force-free flux tube through the corona, say as a cooling effect, how might
the flux tube adjust to find pressure balance again with the local ambient,
bound as it is within the lines of the background magnetic field?

The commonly held view adopted after \cite{Lust+Schluter1954} is that the flux
tube must adjust as by a reconfiguring of its internal magnetic field to give a
more aligned current so that it becomes nearer force-free with a smaller
pressure contrast.  However, in unembedded or embedded conditions, the total
internal transverse and longitudinal magnetic fluxes like the current are
conserved along the flux-tube length.  If the flux-tube magnetic field adjusts
to become nearer force-free, it must do so everywhere along its length, but it
may be prevented from doing so by ongoing footpoint motions originating outside
the coronal atmosphere.

The pressure weakening may thus force the flux tube to change geometrically in
relation to the background magnetic field in the locale of the cooling, against
its embedded-field equilibrium. Small deviations from a stable equilibrium are
self-restoring, with Lorentz forces arising in the vicinity of the flux tube.
The Lorentz forces drive outflows to refill the depleted atmosphere, heating it
to its equilibrium temperature, mainly compressively as the most significant
effect, as outlined in \cite{November+Koutchmy1996} (hereafter NK96), and
\cite{November2004} (hereafter N04).  As the equilibrium atmosphere is restored,
the flux tube is brought back into alignment with the background magnetic field,
with it again satisfying the \cite{Lust+Schluter1954} constraint that its
non-force-free degree lies within the ambient gas-pressure contrast.

Solar spicules may be evidence of equilibrium-restoring flux-tube refilling
outflows.  If spicule outflows are smoothly accelerated into the solar corona,
\cite{Athay+Holzer1982} show that they can carry adequate kinetic energy to heat
the whole solar atmosphere and account for all radiative, conductive, downflow,
and outward wind energy losses.

There seems to be broad consensus that solitary current-carrying channels can
be introduced into a large-scale magnetic field due to isolated footpoint
motions, though the effect of nonlocalized footpoint motions is a matter of
considerable theoretical debate \citep{Zweibel+LiH-S1987, ParkerEN1994}.  There
has been much interest in magnetic footpoint motions in solar and stellar
coronae because accumulating field distortions may eventually lead to unstable
topological configurations and reconnection events, with the partial or complete
reformation of the large-scale background magnetic field, as in a solar coronal
mass ejection.  Here however activity is an aside.  Rather ongoing localized
footpoint motions or the introduction of new magnetic flux from the outside
appears to be able to produce isolated current-carrying fields embedded in a
background near-potential magnetic field as the prevailing state of the system.

Models of magnetic field distortions induced by footpoint motions often allow
general {\it non}-force-free magnetic fields frozen into the flow or embedded in
the background field \citep{Syrovatskii1971, ParkerEN1972,
vanHoven+MaSS+Einaudi1981, ChengCZ+ChoeGS1998, KumarD+Bhattacharyya2011}.
Flux-tube footpoint motions can produce compression and torque that introduce a
non-force-free boundary condition outside the coronal atmosphere as in the solar
photosphere \citep{Karpen+++1993, Abbett2007}.  Photospheric magnetic fields may
be overly constrained by the force-free assumption, with embedded non-force-free
current-carrying fields extending into the corona above \citep{Choe+Jang2013}.
The tendency for flux tubes to relax to force free along their lengths, as seen
in solar simulations \citep{Mikic+Schnack+vanHoven1989}, must be opposed by
ongoing footpoint motions.

By direct pressure balance, non-force-free current-carrying fields embedded in a
large-scale background field are shown to necessitate a minimum surrounding
atmosphere, whose gas pressure is everywhere fixed in proportion to the
background magnetic pressure, as described in Section \ref{s:fluxtube}. The
relative strength of that atmosphere, the constant of proportionality, is
determined by the overall strongest globally distributed non-force-free embedded
fields.

An embedded current-carrying field is represented formally using a
transverse-longitudinal decomposition as elaborated in Section \ref{s:embed}.
With this field specification, the general MHD pressure equation for the overall
equilibrium, including flows and gravity as part of a wide class of physical
processes, separates into independent global and local relations, as discussed
in Section \ref{s:MHs}.  The local pressure relation for the embedded field
takes the form of the Grad-Shafranov (GS) equation, but with 3D component fluxes
and relative pressure contrast, which are conserved as the embedded field
spreads out or changes shape in following the background magnetic field.  Local
equilibrium solutions are described in Section \ref{s:local}.

The global relation reproduces the MHD pressure equation without the embedded
fields, but with coupled ambient gas and magnetic pressures. A coupled gas
pressure in the magnetically dominant coronal regime determines the equilibrium
distribution of density, temperature, and accelerated outflow throughout the
global atmosphere.

Predicted coronal temperatures are corroborated in a sampling of observations of
the sun, stars, and the Milky Way, as described in Section \ref{s:glob}. Many
planetary bodies in the solar system exhibit magnetospheres and show evidence
for embedded flux tubes, which may be being introduced at their magnetopauses by
the solar wind.  They do characteristically also exhibit a thermosphere, that is
a sharp temperature rise like a solar or stellar transition region, and a
surrounding hot exosphere like a corona above, with polar outflowing winds
sometimes present.  If extreme objects, neutron stars and black holes, have
surrounding magnetic fields with embedded non-force-free current-carrying
fields, they are predicted to have exceedingly highly energized coronae.

A widely distributed hot gas for the interstellar background of the Milky Way
has been interpreted as evidence for a galactic corona \citep{Spitzer1956,
Spitzer+JenkinsEB1975}.  Spiral galaxies exhibit large-scale spatially coherent
magnetic fields on the scale of their visible disks suggesting a regime of
magnetic dominance, whereas in elliptical galaxies, magnetic fields of
comparable strength are seen but without a large-scale spatially coherent
structure \citep{MossD+Shukurov1996}. Spiral galaxies show evidence too of much
fine filamentary structure, fountains, which like solar spicules, may feed the
hot interstellar gas, a magnetic-gas co-rotation, and strong outflowing polar
winds. However galaxies have a cosmic-ray pressure comparable to their magnetic
pressure, with very low densities and long collision times, and very different
structural elements, spatial scales, and lifetimes, which must be considered in
their analyses.

\section{Evidence for Non-Force-Free Flux Tubes in the Solar Corona}
\label{s:obs}

Filamentary density contrast of diameter ${\gtrsim}30$ Mm has been recognized
historically in different solar coronal observations as with polar plumes, but
with suggestions of much finer structures \citep{Newkirk1967}.
\cite{Hewish+Wyndham1963} identified fine elongated approximately radial density
contrast smaller than 5 Mm in the solar corona and interplanetary medium around
$60R_\sun$ from radio scintillation measurements.  The scintillation spectrum
extends to much finer scales in the interpretation of multiple baseline
measurements, suggesting an inner scale of turbulence of about 1 km at $5R_\sun$
that increases to about 100 km near $100R_\sun$ \citep{Coles+Harmon1989}.  Such
scintillation observations may suggest radially elongated filamentary cones of
increasing diameter with radial height \citep{WooR+Habbal1997}.  However radio
scintillation measurements represent very-fine-scale spatial-spectral inferences
that seem difficult to reconcile with simultaneous space coronagraph and eclipse
images \citep{Woo2007}.

Theoretical reconstructions from in-situ spacecraft measurements indicate
${\simeq}20$ Mm scale filamentary density variations in the solar wind around 1
AU, some with a non-force-free signature \citep{HuQ+Sonnerup2001,
Romashets+Vandas2005}.

Nonuniform accelerations in comet tails have suggested fine filamentary coronal
structure as reviewed historically \citep{Antrack+Biermann+Lust1964}.  In 2011,
comet Lovejoy came as close as 0.2 $R_\sun$ above the solar photosphere, and
showed intermittent tail accelerations tracing a spatial signature of 4 Mm
filamentary density structure, which appears to be aligned with the background
magnetic field \citep{McCauley+++2013, Downs+++2013}.

Dark filamentary `voids' are seen against a fairly uniform coronal background in
white-light eclipse observations \citep{Koutchmy+Laffineur1970,
Rusin+Rybansky1985}, and in space-based coronagraph images
\citep{MacQueen+Sime+Picat1983}. At the unique 1991 total eclipse over Mauna
Kea, the 3.6m Canada-France-Hawaii Telescope (CFHT) obtained a high-resolution
white-light photographic time series showing both dark and bright thread-like
filaments as fine as the seeing resolution limit of about 0.7 Mm, but with
maximum power ranging up to about 5 Mm (NK96).  Since that observation,
beautiful quality photoelectric white-light eclipse images have been obtained,
which show remarkable fine structure as bright polar plumes, and both dark and
bright fine filaments tracing out the lines of the background magnetic field
(see \citealt{Koutchmy+++2007, Rusin+++2008, Rusin+++2010, Ambroz+++2009,
Pasachoff+++2007, Pasachoff+++2009, Pasachoff+++2011}; and contained references
and citations).

Polarized white-light intensity is produced by Thomson scattering and thus
proportional to electron density or density in the highly ionized coronal
plasma.  The observed range of density contrast from evacuated to twice the
ambient in the coronal filaments (NK96) is the predicted range for static
non-force-free flux tubes in a relatively high-beta coronal atmosphere, as
elaborated in N04 and in Section \ref{s:fluxtube}.

Only an unreasonable temperature many times larger than the ambient could
produce the observed degree of evacuation seen in the dark filaments, but
filaments, most obviously the bright ones, exhibit a uniform density contrast
with the background, indicating a hydrostatic temperature that is close to the
ambient.  Waves have been suggested, but the fine filamentary structure is quite
persistent over the 4 minute eclipse time span in NK96.  Rather, feature changes
between eclipse images obtained 90 minutes apart from two sites along the
eclipse path appear to suggest only a slow evolution in the large-scale coronal
structure (\citealt{Vsekhsviatskii+++1975}; \citealt{Koutchmy1999a}, Figure 7).

Narrow steady jet flows, which must follow magnetic field lines by the frozen-in
condition for the nearly fully conductive coronal plasma, add a Bernoulli
displacement pressure that offsets the gas pressure and may thus be able to
produce dark filaments, but cannot simply explain the overdense bright
filaments.  Flows might contribute to the displacement pressure significantly
only if they are a good fraction of the Alfv{\'e}n speed, but there appears to
be no direct observational evidence for flows in fine coronal filaments, as have
long been seen following inhomogeneities as tracers in quiescent prominences on
the limb (see \citealt{Tandberg-Hanssen1974}, \S2.23 and \S2.26 and references,
\citealt{Darvann+Koutchmy+Zirker1989}), and on the disk in H$\alpha$
\citep{LinY+Engvold+Wiik2003}.  Possible consequences of high-speed flows such
as field distortions or fluid turbulence do not seem to be detectable either in
the lower quiescent corona, although the CFHT eclipse observation set includes
that remarkable exceptional case of a single outward moving ${<}1$ Mm plasmoid
\citep{Vial+++1992}.

\section{Direct Flux-Tube Pressure Balance}
\label{s:fluxtube}

\begin{figure}[ht!]
\centering\noindent
\centerline{\includegraphics[width=0.7\columnwidth,bb=59 150 730 470]{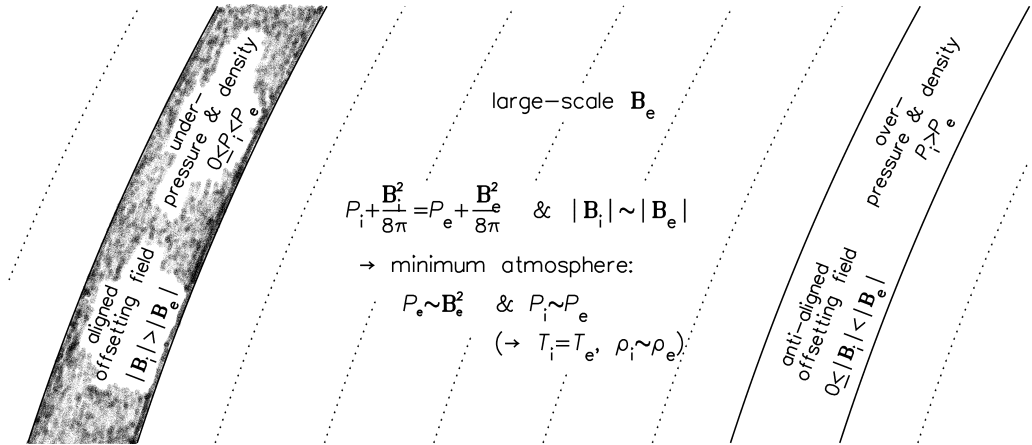}} 
\caption{Embedded non-force-free underdense (dark) and overdense (bright) flux
tubes running parallel to the large-scale background magnetic field (dotted
lines).  Direct pressure balance requires a minimum ambient atmosphere outside
evacuated flux tubes determined solely by the background field
$P_{\rm{e}}\sim{\bf{B}}_{\rm{e}}^2$, which gives a flux-tube pressure and
density proportional to the ambient $P_{\rm{i}}\sim P_{\rm{e}}$,
$\rho_{\rm{i}}\sim\rho_{\rm{e}}$, and temperature equal to it
$T_{\rm{i}}=T_{\rm{e}}$.}
\label{f:ftcor}
\end{figure}

Idealized non-force-free flux tubes embedded in a large-scale background
magnetic field satisfy the basic pressure-balance relations illustrated in
Figure \ref{f:ftcor} and described in this section.  A flux tube of elevated
internal magnetic field strength $|{\bf{B}}_{\rm{i}}|$ due to a parallel and
aligned offsetting field displaces the external ambient gas pressure
$P_{\rm{e}}$ producing an internal underpressure $P_{\rm{i}}$, or evacuation in
the limiting case, illustrated on the left in the figure.  A flux tube of
diminished magnetic field strength $|{\bf{B}}_{\rm{i}}|$ due to a parallel but
{\it anti}-aligned offsetting field exhibits an internal overpressure
$P_{\rm{i}}$, illustrated on the right.  The internal and external total
gas-plus-magnetic pressures must balance
\begin{equation}
P_{\rm{i}}+{{\bf{B}}_{\rm{i}}^2\over8\pi}=
P_{\rm{e}}+{{\bf{B}}_{\rm{e}}^2\over8\pi},
\label{e:Pdif}
\end{equation}
as is justified for current-carrying fields embedded in a background potential
or non-potential magnetic field as elaborated in Section \ref{s:local}, written
in Gaussian units for flux tubes without interal flows and ignoring small
buoyancy effects due to any possible temperature differences with the ambient.

For a small-scale magnetic field superposed on a stronger background field,
\cite{ParkerEN1972} shows that equilibrium is possible only if the pattern of
small-scale variations is uniform along the large-scale background field with
${\bf{B}}_{\rm{i}}\parallel{\bf{B}}_{\rm{e}}$.  Field parallelism is consistent
with coronal observations, which show fine filaments following the background
field and tracing out its large-scale organization over their lengths (e.g.\
NK96, or see \citealt{Sakurai+Uchida1977}).  Flux conservation over the cross
section of the embedded field then gives an internal field strength proportional
to the external field strength
\begin{equation}
{\bf{B}}_{\rm{i}}^2=(1+\beta){\bf{B}}_{\rm{e}}^2,
\label{e:Bpar}
\end{equation}
for $\beta$ constant along the embedded field, with $\beta>0$ for aligned and
$\beta<0$ for anti-aligned offsetting embedded fields. Strictly, the
\cite{ParkerEN1972} parallelism of the embedded field is for an idealized
straight background field.  Limitations in the accuracy of the embedded-field
parallelism are elaborated at the end of Section \ref{s:embed} and in Section
\ref{s:local}.

Using Eq.\ (\ref{e:Bpar}) to eliminate ${\bf{B}}_{\rm{i}}^2$ from Eq.\
(\ref{e:Pdif}) gives
\begin{equation}
P_{\rm{e}}=\beta{{\bf{B}}_{\rm{e}}^2\over8\pi}+P_{\rm{i}}.
\label{e:Patm1}
\end{equation}
Evacuation $P_{\rm{i}}=0$ in the strongest aligned embedded fields of largest
$\beta$ seems natural as Lorentz-force-driven outflows cease when the
equilibrium condition Eq.\ (\ref{e:Pdif}) is minimally met. Thus the atmospheric
pressure relation may be written
\begin{equation}
P_{\rm{e}}=\beta_{\rm{c}}{{\bf{B}}_{\rm{e}}^2\over8\pi},
\label{e:Patm2}
\end{equation}
with the pressure coefficient $\beta_{\rm{c}}=\beta$ for the overall strongest
embedded fields over the globe.  The gas pressure $P_{\rm{e}}$ for the minimum
atmosphere everywhere outside embedded fields is proportional to the background
magnetic pressure ${\bf{B}}_{\rm{e}}^2/(8\pi)$.

In an atmosphere $P_{\rm{e}}$ of given $\beta_{\rm{c}}$ from Eq.\
(\ref{e:Patm2}), embedded fields with aligned offsetting fields of weaker excess
field strength $\beta<\beta_{\rm{c}}$ in Eq.\ (\ref{e:Patm1}) exhibit the
positive internal pressure
\begin{equation}
P_{\rm{i}}=(1-{\beta\over\beta_{\rm{c}}})P_{\rm{e}},
\label{e:Patm3}
\end{equation}
so as the embedded field ranges in strength from $0<\beta\leq\beta_{\rm{c}}$,
its internal pressure goes correspondingly from $P_{\rm{e}}>P_{\rm{i}}\geq0$.

An embedded field with an anti-aligned offsetting magnetic field, as on the
right in the drawing in Figure \ref{f:ftcor}, must exhibit a reduced internal
field strength ranging from $0\leq{\bf{B}}_{\rm{i}}^2<{\bf{B}}_{\rm{e}}^2$.
Even though anti-aligned flux tubes may in principle exhibit a net reversed
internal field with ${\bf{B}}_{\rm{i}}$ directed oppositely to
${\bf{B}}_{\rm{e}}$, the magnetic displacement pressure is always nonnegative
with the minimum ${\bf{B}}_{\rm{i}}^2=0$.  The realizable range of magnetic
field strengths from $0\leq{\bf{B}}_{\rm{i}}^2<{\bf{B}}_{\rm{e}}^2$ corresponds
to the range $-1\leq\beta<0$ in Eq.\ (\ref{e:Bpar}), giving the corresponding
range of internal pressures $(1+1/\beta_{\rm{c}})P_{\rm{e}}\geq
P_{\rm{i}}>P_{\rm{e}}$ from Eq.\ (\ref{e:Patm3}).

The full range of internal gas pressure for coronal flux tubes is from $0\leq
P_{\rm{i}}\leq(1+1/\beta_{\rm{c}})P_{\rm{e}}$ as their field strengths range
from $\beta_{\rm{c}}\geq\beta\geq-1$.  With their internal pressures
everywhere proportional to the external ambient from Eq.\ (\ref{e:Patm3}), their
hydrostatic temperatures must be the same as the local ambient
$T_{\rm{i}}=T_{\rm{e}}$ from the hydrostatic relation, i.e.\ Eq.\
(\ref{e:corHT2}) in Section \ref{s:glob}.  Then by the perfect gas law, their
internal densities must be proportional $\rho_{\rm{i}}\sim\rho_{\rm{e}}$.  In
the limiting strongest stable atmosphere $\beta_{\rm{c}}=1$, flux tubes range
in density from evacuated to two times the ambient, which is what is seen in the
solar corona shown in Figures 14 and 15 of NK96.

With a global distribution of flux tubes of similar strength, there will only be
one atmospheric coupling constant $\beta_{\rm{c}}$ for the large-scale global
atmosphere. The strength of the flux tubes might really be fixed by the limit
for global magnetic stability rather than by any particular property intrinsic
to flux tubes.  Active-region magnetic fields are very much stronger than
quiet-sun fields but of small global surface areal coverage.  Energized only by
local subsurface convective motions, they probably cannot raise the whole
surrounding coronal atmosphere up to their magnetic pressures, and so must tend
to relax to near force free for pressure balance with the surrounding quiet
corona. Comparisons between solar coronal density and extrapolated magnetic
field strength do suggest that the solar corona has a gas pressure comparable to
the magnetic pressure with $\beta_{\rm{c}}\simeq1$ in Eq.\ (\ref{e:Patm2})
outside active regions \citep{GaryGA2001}.

Active-region temporally varying or average components are known to contribute
only a small fraction to the total solar coronal X-ray emission
\citep{Orlando+Peres+Reale2004, Argiroffi+++2008}.  However it seems possible to
include their physics in an overall coronal picture.  Solar active regions
appear to consist mainly of closed low-lying field lines with a background
magnetic field that is very much locally enhanced and separated, even possibly
abruptly so, from the surrounding quiet corona.  The active-region background
field contains its own embedded current-carrying fields, which must be near
force free and described by the same pressure balance Eq.\ (\ref{e:Pdif}), but
with gas pressures inside and outside $P_{\rm{iar}}$ and $P_{\rm{e}}$ small
compared to the magnetic pressures ${\bf{B}}_{\rm{iar}}^2/(8\pi)$ and
${\bf{B}}_{\rm{ear}}^2/(8\pi)$, that is with
${\bf{B}}_{\rm{iar}}^2\simeq{\bf{B}}_{\rm{ear}}^2$ or $|\beta|\ll1$ in Eq.\
(\ref{e:Bpar}).  Active-region current-carrying fields may reach evacuation
$P_{\rm{iar}}=0$ even though they have an internal magnetic field strength only
a little greater than that of their embedding magnetic field
${\bf{B}}_{\rm{iar}}^2\gtrsim{\bf{B}}_{\rm{ear}}^2$, giving a coronal beta
$\beta_{\rm{car}}$ that is very small in Eq.\ (\ref{e:Patm2}), distinctly less
than that of the quiet sun $\beta_{\rm{car}}\ll\beta_{\rm{c}}\lesssim1$.

\section{Embedded Form}
\label{s:embed}

As a more comprehensive approach, the localized current-carrying magnetic field
is decomposed into transverse and longitudinal components from a main vector
axis as
\begin{equation}
{\bf{B}}={\bf\nabla}{\bf\times}(\check{a}{\bf\Omega})+\check{b}{\bf\Omega},
\label{e:B0}
\end{equation}
where the functions marked by an over-check are called the `constituent
functions' of the embedded field, which vary only in its vicinity. For a
localized embedded field, the constituent functions asymptotically become
$\check{a}\rightarrow0$ and $\check{b}\rightarrow\check{b}_{\rm{e}}$ a constant
far from the embedded field.

The vector axis ${\bf\Omega}$ for the magnetic field is commonly taken as a
constant coordinate direction ${\bf\Omega}\rightarrow{\bf\nabla}z$, which leads
to the well-known 2D MHD GS solutions developed in the literature in different
ways in a number of publications \citep{Lust+Schluter1954, Lust+Schluter1957,
Chandrasekhar1956a, Grad+Rubin1958, Shafranov1958}.  Greater flexibility is
allowed by taking the vector axis ${\bf\Omega}$ to be a general Euler coordinate
or flux surface for the magnetic field (e.g.\ \citealt{D'haeseleer+++1991}).
With the continuity of the magnetic field
\begin{equation}
{\bf\nabla}{\bf\cdot}{\bf{B}}=0,
\label{e:divB}
\end{equation}
it follows
\begin{equation}
{\bf\nabla}{\bf\cdot}(\check{b}{\bf\Omega})=0,
\label{e:divb0}
\end{equation}
and applying Gauss' theorem, the longitudinal magnetic flux, the cross-sectional
areal integral of $\check{b}{\bf\Omega}$, must be conserved over every
perpendicular surface within any given set of field lines.  Taking the flux
surface ${\bf\Omega}$ to be a conservative vector field
\begin{equation}
{\bf\nabla}{\bf\cdot}{\bf\Omega}=0,
\nonumber
\end{equation}
consistent with its usage as an Euler coordinate, then gives a conserved
longitudinal constituent function $\check{b}$ along the flux surface
${\bf\Omega}$
\begin{equation}
{\bf\Omega}{\bf\cdot}{\bf\nabla}\check{b}=0.
\label{e:divb1}
\end{equation}

The transverse magnetic flux is taken to be similarly conserved
\begin{equation}
{\bf\nabla}{\bf\cdot}(\check{a}{\bf\Omega})=0,
{\rm\ giving}\quad {\bf\Omega}{\bf\cdot}{\bf\nabla}\check{a}=0,
\label{e:diva}
\end{equation}
for a constituent function $\check{a}$ constant along flux-surface lines.  The
transverse constraint is a gauge choice, the Coulomb gauge, which simplifies the
current density ${\bf{J}}$ and Lorentz force ${\bf{F}}$.  There appears to be no
possibility for finding explicit isotropic solutions otherwise, as many dot
products ${\bf\Omega}{\bf\cdot}{\bf\nabla}\check{a}$ arise in independent ways
in expansions of the Lorentz force.  The magnetic-field transverse and
longitudinal {\it fluxes} are conserved, that is the areal integrals of
$\check{a}{\bf\Omega}$ and $\check{b}{\bf\Omega}$ over every cross section
within any given set of field lines, whereas the constituent function remapped
{\it values} $\check{a}$ and $\check{b}$ remain constant even as the field
spreads out or changes shape.

The current density ${\bf{J}}=\sfrac{c}{4\pi}{\bf\nabla}{\bf\times}{\bf{B}}$
with a conserved transverse constituent function $\check{a}$ from Eq.\
(\ref{e:diva}) is written
\begin{equation}
{\bf{J}}={c\over4\pi}\left({\bf\nabla}{\bf\times}(\check{b}{\bf\Omega})
-{\bf\nabla}^2(\check{a}{\bf\Omega})\right),
\label{e:J0}
\end{equation}
leading to a Lorentz force ${\bf{F}}=\sfrac{1}{c}{\bf{J}}{\bf\times}{\bf{B}}$,
which expands into four cross-product terms
\begin{equation}\begin{split}
&{\bf{F}}={1\over4\pi}\biggl(
\bigl\{({\bf\nabla}{\bf\times}(\check{b}{\bf\Omega})){\bf\times}({\bf\nabla}{\bf\times}(\check{a}{\bf\Omega}))\bigr\}_1
+\bigl\{({\bf\nabla}{\bf\times}(\check{b}{\bf\Omega})){\bf\times}\check{b}{\bf\Omega}\bigr\}_2\\
&-\bigl\{({\bf\nabla}^2(\check{a}{\bf\Omega})){\bf\times}({\bf\nabla}{\bf\times}(\check{a}{\bf\Omega}))\bigr\}_3
-\bigl\{({\bf\nabla}^2(\check{a}{\bf\Omega})){\bf\times}{\bf\nabla}(\check{b}{\bf\Omega})\bigr\}_4\biggr).
\label{e:F}
\end{split}\end{equation}
Terms in the expression are numbered by subscripts for later identification.

It is convenient to write the Lorentz force as a sum of three parts with
distinct physical properties, which are studied separately, as
\begin{equation}\begin{split}
{\bf{F}}={\bf{F}}_{\rm{e}}+{\bf{F}}_{\rm{i0}}+{\bf{F}}_{\rm{i1}},
\label{e:Fa}
\end{split}\end{equation}
a global part ${\bf{F}}_{\rm{e}}$ that does not die off away from the embedded
field, and local parts ${\bf{F}}_{\rm{i0}}$ and ${\bf{F}}_{\rm{i1}}$ that do.
All of the Lorentz-force terms in the expansion in Eq.\ (\ref{e:F}) that contain
the constituent function $\check{a}$ or a derivative of $\check{b}$ do die off
and are local, which leaves only one vector cross product out of the second term
that does not die off locally, written
\begin{equation}
{\bf{F}}_{\rm{e}}={\check{b}^2\over4\pi}
({\bf\nabla}{\bf\times}{\bf\Omega}){\bf\times}{\bf\Omega}.
\label{e:Fe0}
\end{equation}
With the flux surface ${\bf\Omega}$ becoming asymptotically parallel to the
background field as it must in Eq.\ (\ref{e:B0}),
\begin{equation}
{\bf{B}}_{\rm{e}}=\check{b}_{\rm{e}}{\bf\Omega},
\label{e:Blim}
\end{equation}
and the global Lorentz force becomes consistently the Lorentz force due to a
background non-potential magnetic field ${\bf{B}}_{\rm{e}}$
\begin{equation}
{\bf{F}}_{\rm{e}}={1\over4\pi}
({\bf\nabla}{\bf\times}{\bf{B}}_{\rm{e}}){\bf\times}{\bf{B}}_{\rm{e}}.
\label{e:Fe1}
\end{equation}

The remaining terms in the expanded Lorentz force are grouped into a part
${\bf{F}}_{\rm{i0}}$ where the flux surface ${\bf\Omega}$ appears
undifferentiated, and into a part ${\bf{F}}_{\rm{i1}}$ otherwise.  As long as
the scale of variation of the flux surface $H_{\Omega}=H_{\rm{Be}}$ is large
compared to a characteristic scale of variation in the current-carrying field
$d$, the derivatives of the basis vector will be relatively small compared to
those of the constituent functions, giving ${\bf{F}}_{\rm{i1}}$ of relative
order $d/H_{\rm{Be}}$ compared to ${\bf{F}}_{\rm{i0}}$.  The
more-significant-order Lorentz-force part ${\bf{F}}_{\rm{i0}}$ is expanded
\begin{equation}
{\bf{F}}_{\rm{i0}}={{\bf\Omega}^2\over4\pi}\biggl(
\bigl\{{\bf\nabla}\check{b}{\bf\times}{\bf\nabla}\check{a}\bigr\}_1
-\bigl\{\check{b}{\bf\nabla}\check{b}\bigr\}_2
-\bigl\{{\bf\nabla}^2\check{a}{\bf\nabla}\check{a}\bigr\}_3\biggr),
\label{e:Fi0a}
\end{equation}
applying vector identities to the respective terms in Eq.\ (\ref{e:F})
straightforwardly, with the fourth term disappearing as it reduces to a cross
product between two vectors both parallel to ${\bf\Omega}$.  The usual
notational convention is adopted that differential operators, as gradient
${\bf\nabla}$, Laplacian ${\bf\nabla}^2$, or tensorial
$\partial_{\imath}={\bf\nabla}_{\imath}$, have precedence over multiplication,
operating only on their first following variable.

The MHD approximation with an isotropic pressure is commonly taken to be
applicable even in very low-density astrophysical conditions, as mixing
processes, especially Coulomb scattering, tend to isotropize the gas pressure
still on scales small compared to the magnetic structure.  The equilibrium
magnetic flux is thus expected to have a spatial distribution with minimal
anisotropy in its Lorentz force. In ideal cases, like the static GS straight
field geometry, isotropic pressure solutions are found, but anisotropic effects
cannot be avoided in more usual plasma problems (e.g.\ \citealt{Bellan2008,
Biskamp2008}). The external factor ${\bf\Omega}^2$ in the expression for the
local Lorentz force in Eq.\ (\ref{e:Fi0a}) represents a slow global variation
that becomes identified with external forces in the separation of the gas
pressure into globally and locally varying relations, as described in Section
\ref{s:MHs} and Appendix \ref{a:MHs1}.  For local isotropy, the ratio
${\bf{F}}/{\bf\Omega}^2$ is left equated to the gradient of a scalar pressure.

A Lorentz force that is locally isotropic with
${\bf\nabla}{\bf\times}({\bf{F}}/{\bf\Omega}^2)=0$ is sought, at least as a weak
constraint that may not always be possible to satisfy exactly.  The second term
in Eq.\ (\ref{e:Fi0a}) always vanishes under the curl
${\bf\nabla}{\bf\times}({\check{b}}{\bf\nabla}{\check{b}})=
{\bf\nabla}{\bf\times}({\bf\nabla}{\check{b}}^2)/2=0$. The vanishing of the
third term under the curl
${\bf\nabla}{\bf\times}({\bf\nabla}^2\check{a}{\bf\nabla}\check{a})=
{\bf\nabla}({\bf\nabla}^2\check{a}){\bf\times}{\bf\nabla}\check{a}=0$, requires
that the vectors be everywhere parallel
${\bf\nabla}\check{a}\parallel{\bf\nabla}({\bf\nabla}^2\check{a})$.  The three
underlying Jacobian $2\times2$ determinants in the cross product are zero (with
at least one nontrivially so), implying the functional dependency
${\bf\nabla}^2\check{a}=({\bf\nabla}^2\check{a})[\check{a}]$ or the inverse
$\check{a}=\check{a}[{\bf\nabla}^2\check{a}]$, which may be multivalued
dependencies.  For clarity in the subtle functional relations, square braces are
used throughout to denote a functional dependence.

As both $\check{a}$ and $\check{b}$ are constant in the direction of the flux
surface ${\bf\Omega}$ by Eqs.\ (\ref{e:diva}) and (\ref{e:divb1}), their
gradients are both perpendicular to ${\bf\Omega}$, and the cross product, the
first term in Eq.\ (\ref{e:Fi0a}), parallel
${\bf\nabla}\check{a}{\bf\times}{\bf\nabla}\check{b}\parallel{\bf\Omega}$, so
the cross product is unidirectional locally with
${\bf\Omega}\parallel{\bf\nabla}z$.  For the curl of the cross product to
vanish, the term itself must be the gradient of a scalar function as
${\bf\nabla}\check{a}{\bf\times}{\bf\nabla}\check{b}={\bf\nabla}f[z]=(df/dz){\bf\nabla}z$
for $f[z]$ a function of a single coordinate $z$.  The cross product
${\bf\nabla}\check{a}{\bf\times}{\bf\nabla}\check{b}=(df/dz){\bf\nabla}z$ is a
function of $z$ alone, but as the functions $\check{a}$ and $\check{b}$ are
bounded in their variations away from the center of the embedded field, they
must exhibit functional dependencies perpendicular to ${\bf\nabla}z$, as so must
any nonzero cross product ${\bf\nabla}\check{a}{\bf\times}{\bf\nabla}\check{b}$.
Variations perpendicular to $z$ in the cross product contradict its
unidirectional requirement, so bounded solutions are only possible with the
vanishing of the cross product itself everywhere
${\bf\nabla}\check{a}{\bf\times}{\bf\nabla}\check{b}=0$, which necessitates the
multivalued functional dependencies $\check{b}=\check{b}[\check{a}]$ or the
inverse $\check{a}=\check{a}[\check{b}]$.

The continuity of $\check{a}$ in Eq.\ (\ref{e:diva}) and $\check{b}$ in Eq.\
(\ref{e:divb1}) are consistent with, though not necessarily implying, the
functional constraint $\check{b}=\check{b}[\check{a}]$.  Without the continuity
of $\check{a}$ as a gauge choice, the functional constraint
$\check{b}=\check{b}[\check{a}]$ cannot be satisfied.  The form of the solution
described here is only consistent with the choice of gauge.

Applying the functional constraints
${\bf\nabla}^2\check{a}=({\bf\nabla}^2\check{a})[\check{a}]$ and
$\check{b}=\check{b}[\check{a}]$ to Eq.\ (\ref{e:Fi0a}) gives
\begin{equation}
{\bf{F}}_{\rm{i0}}=-{{\bf\Omega}^2\over4\pi}
\left(({\bf\nabla}^2\check{a})[\check{a}]+{1\over2}{d\check{b}^2\over d\check{a}}\right){\bf\nabla}\check{a},
\label{e:Fi0b}
\end{equation}
which resembles the classical 2D GS Lorentz force for straight geometries, but
now written for 3D Laplacian and constituent functions.

The remaining smaller-order local Lorentz-force ${\bf{F}}_{\rm{i1}}$ reduces as
elaborated in Appendix \ref{a:Fi}.  It vanishes in a potential flux surface
${\bf\Omega}$ in Cartesian current-sheet solutions to give the purely isotropic
Lorentz force written in Eq.\ (\ref{e:Fi0b}).  Otherwise a small-order
anisotropic Lorentz force $F_{\rm{i1}}$ of order
${\bf{B}}_{\rm{e}}^2/H_{\rm{Be}}$ arises.  Compared to the isotropic Lorentz
force $F_{\rm{i0}}$ from Eq.\ (\ref{e:Fi0b}), which is of order
${\bf{B}}_{\rm{e}}^2/d$, the anisotropic Lorentz force has the small relative
order $d/H_{\rm{Be}}$, for $d$ the main scale of variation in the
current-carrying field.

With a vanishing transverse field in the current-carrying field away from its
center, asymptotic parallelism of the embedded and background fields away from
the embedded field, as in Eq.\ (\ref{e:Blim}), is an implicit feature of the
magnetic-field specification Eq.\ (\ref{e:B0}).  However that field
specification is not the most general possible.  A vector field can be added to
Eq.\ (\ref{e:B0}) as a third independent degree of freedom, as with the addition
of a potential field in a Helmholtz decomposition (\citealt{Morse+Feshbach1953},
Section 13.1; \citealt{ParkerEN1979}, p.\ 542).  It is sometimes possible to
redefine the terms in the superposition to give back the form for the magnetic
field in Eq.\ (\ref{e:B0}).  Otherwise the background magnetic field will not be
parallel to the flux surface ${\bf{B}}_{\rm{e}}\not\parallel{\bf\Omega}$, but
then new anisotropic terms are also added to the Lorentz force of
most-significant order, as is straightforwardly demonstrated.  These lead to
anisotropic Lorentz forces of relative order the angular deviation between the
flux surface and background field $d/H_{\rm{Be}}$.

The scale height for a solar potential spherical-harmonic of degree $\ell=2$ is
$H_{\rm{Be}}=R_\sun/(\ell+2)=174$ Mm.  With flux tubes of diameter
$d_{\rm{cc}}\simeq0.7$ Mm from the NK96 eclipse observation,
$d_{\rm{cc}}/H_{\rm{Be}}\simeq4$E$-3$, with still finer flux tubes likely
present in the lower corona as are inferred farther out.  In thin-sheathed flux
tubes embedded in a background potential field, which are described in Section
\ref{s:local}, the characteristic scale of variation is the sheath thickness
$d=d_{\rm{sh}}$, which may be much smaller than the flux-tube diameter.  The
anisotropic Lorentz forces are correspondingly smaller by the factor
$d_{\rm{sh}}/d_{\rm{cc}}$, and confined to the sheath.

\section{Solution Separability}
\label{s:MHs}

This work develops an overall equilibrium solution strategy for embedded
current-carrying fields based upon the separation of the MHD equations into
independent global and local relations like what is advanced in N04, but now
shown to be applicable in more wide-ranging circumstances, as with flows and
temperature variations.  The MHD pressure balance is expressed
\begin{equation}
{\bf\nabla}P+{P\over H_{\rm{s}}g[R]}{D{\bf{u}}\over Dt}+{P\over H_{\rm{s}}}{\bf\nabla}R=
{1\over4\pi}({\bf\nabla}{\bf\times}{\bf{B}}){\bf\times}{\bf{B}},
\label{e:MHs1}
\end{equation}
with the gas pressure $P$, static pressure scale height $H_{\rm{s}}$, flow
vector ${\bf{u}}$, inward gravitational acceleration $g[R]$, with the radial
coordinate $R$ from the gravitating center, and total time derivative for the
flow field $D{\bf{u}}/Dt=\partial{\bf{u}}/\partial
t+{\bf{u}}{\bf\cdot}{\bf\nabla}{\bf{u}}$. Time derivatives are retained in the
global ambient equation to explore near equilibrium effects, but the steady
equilibrium with zero time derivatives is of principle interest.  The density
$\rho$ is eliminated in favor of the static pressure scale height $H_{\rm{s}}$
with $P/H_{\rm{s}}=\rho g[R]$.  Supposing a perfect gas with $P=\rho kT/\bar{m}$
gives $H_{\rm{s}}=kT/(\bar{m}g[R])$ proportional to the temperature $T$, with
Boltzmann's constant $k$, and the mean particle mass $\bar{m}$.

The equations are completed with the continuity of the flow field and the
magnetic-induction equation for a fully conductive plasma
\begin{equation} 
{\partial\rho\over\partial t}+{\bf\nabla}{\bf\cdot}(\rho{\bf{u}})=0,
\quad\quad{\partial{\bf{B}}\over\partial t}={\bf\nabla}{\bf\times}\left({\bf{u}}{\bf\times}{\bf{B}}\right).
\label{e:ucont}
\end{equation}
Upper-case symbols $(R,\Theta,\Phi)$ are used throughout to denote the global
spherical large-scale, e.g.\ helio-, stellar-, planetary-, or galactic-centered
system, a convention common in astrophysics.  Lower-case symbols are reserved
for the local coordinates of the embedded field, as Cartesian $(x,y,z)$, or
cylindrical $(r,\phi,z)$.

In the present section, the background magnetic field is assumed to be
potential, and the static scale height $H_{\rm{s}}$ or temperature $T$, and flow
field ${\bf{u}}$ taken to be varying globally without showing any systematic
variations near embedded fields.  Effects of a non-potential background field,
locally changing temperatures and flows, radiation and cosmic-ray pressures, or
viscosity are of interest in this work.  A general strategy for including other
physical processes is discussed at the end of this section and developed in
Appendix \ref{a:MHs1}.

Multiple terms in the MHD pressure balance Eq.\ (\ref{e:MHs1}) are encorporated
into a single integrating factor $\Pi$ with the substitution
\begin{equation}
P=\check{P}\exp\left[-\Pi\right],
\label{e:Pidefa}
\end{equation}
introducing the relative gas pressure $\check{P}$, which varies near embedded
fields, but is constant everywhere outside, and with $\Pi$ varying only
globally.  Taking the gradient leads to
\begin{equation}
{\bf\nabla}P+P{\bf\nabla}\Pi=\exp\left[-\Pi\right]{\bf\nabla}\check{P},
\label{e:Pidefb}
\end{equation}
which is a perfect template matching in its left and right sides those of the
MHD Eq.\ (\ref{e:MHs1}).  The scalar integrating factor $\Pi$ acts as an
isotropic pressure whose gradient must balance the total of all the vector
forces in the global field outside embedded fields.

Laying the template Eq.\ (\ref{e:Pidefb}) over the MHD Eq.\ (\ref{e:MHs1}) gives
for the integrating factor from the equated left-sides
\begin{equation}
{\bf\nabla}\Pi={1\over H_{\rm{s}}g[R]}{D{\bf{u}}\over Dt}+{1\over H_{\rm{s}}}{\bf\nabla}R,
\label{e:MHs2a}
\end{equation}
and the pressure relation from the equated right sides
\begin{equation}
\exp[-\Pi]{\bf\nabla}\check{P}=
{1\over4\pi}({\bf\nabla}{\bf\times}{\bf{B}}){\bf\times}{\bf{B}},
\label{e:MHs2b}
\end{equation}
which is expanded
\begin{equation}
\exp[-\Pi]{\bf\nabla}\check{P}=
-{{\bf\Omega}^2\over 4\pi}
\left\{({\bf\nabla}^2\check{a})[\check{a}]{\bf\nabla}\check{a}
+{1\over2}{d\check{b}^2\over d\check{a}}{\bf\nabla}\check{a}\right\},
\label{e:MHs2c}
\end{equation}
for the magnetic field from Eq.\ (\ref{e:B0}) using the Lorentz force
${\bf{F}}={\bf{F}}_{\rm{i0}}$ from Eq.\ (\ref{e:Fi0b}), taking for simplicity
the global and small-order parts ${\bf{F}}_{\rm{e}}={\bf{F}}_{\rm{i1}}=0$,
relevant for the assumed potential background magnetic field.  Even in a
potential background field, additional small terms are introduced into the
locally varying part in curly braces with a nonvanishing ${\bf{F}}_{\rm{i1}}$ as
in non-Cartesian solutions, or with ${\bf{F}}_{\rm{e}}$ as for small
potential-field deviations in local current-carrying-field solutions as
elaborated in Section \ref{s:local}.  In a non-potential background magnetic
field, with flows, or with temperature variations, additional principle-order
effects enter into the locally varying part.  Some elaboration of additional
effects are described at the end of this section and in Appendix \ref{a:MHs1}.

In Eq.\ (\ref{e:MHs2c}), all the terms in the integrating factor $\Pi$ and the
flux surface ${\bf\Omega}^2$ vary globally away from and continuously across
embedded fields, whereas the gradient of the relative gas pressure
${\bf\nabla}\check{P}$ and all the terms in curly braces on the right side are
local to the embedded field.  Thus the separation is suggested
\begin{equation}
{\bf\nabla}\check{P}=
-{1\over4\pi}\left(({\bf\nabla}^2\check{a})[\check{a}]+{1\over2}{d\check{b}^2\over d\check{a}}\right){\bf\nabla}\check{a},
\label{e:MHs3a}
\end{equation}
 for the local field, and
\begin{equation}
\exp[-\Pi]={\bf\Omega}^2,
\label{e:MHs3b}
\end{equation}
for the global field.  An arbitrary separation constant is introduced, which has
allowed the factor $4\pi$ to be moved into Eq.\ (\ref{e:MHs3a}) to give the
local relation the exact 2D GS equation form in Gaussian units.

The global relation Eq.\ (\ref{e:MHs3b}) is recast into a more familiar form as
an MHD pressure relation just for the global field.  Taking the natural log of
both sides and operating with the gradient gives
\begin{equation}
{\bf\nabla}\Pi=-{\bf\nabla}\ln[{\bf\Omega}^2],
\nonumber
\end{equation}
then substituting the integrating factor from Eq.\ (\ref{e:MHs2a}) leaves
\begin{equation}
{\bf\nabla}\ln[{\bf\Omega}^2]+{1\over H_{\rm{s}}g[R]}{D{\bf{u}}\over Dt}+{1\over H_{\rm{s}}}{\bf\nabla}R=0.
\label{e:MHs4a}
\end{equation}
A given global distribution for the flux surface ${\bf\Omega}^2$ determines the
static scale height $H_{\rm{s}}$ or temperature, and flow field ${\bf{u}}$. The
relation is indistinguishable from the MHD pressure equation for a global
potential magnetic field.  Substituting with the ambient gas pressure defined in
terms of the flux surface
$P_{\rm{e}}=\beta_{\rm{c}}\check{b}_{\rm{e}}^2{\bf\Omega}^2/(8\pi)$, and taking
the product $\beta_{\rm{c}}\check{b}_{\rm{e}}^2$ to be a global constant gives
\begin{equation} 
{\bf\nabla}P_{\rm{e}}+{P_{\rm{e}}\over H_{\rm{s}}g[R]}{D{\bf{u}}\over Dt}+{P_{\rm{e}}\over H_{\rm{s}}}{\bf\nabla}R=0,
\quad{\rm{with}}\quad
P_{\rm{e}}=\beta_{\rm{c}}{{\bf{B}}_{\rm{e}}^2\over 8\pi},
\label{e:MHs4b}
\end{equation}
equating the flux surface to the ambient magnetic field away from embedded
fields ${\bf\Omega}={\bf{B}}_{\rm{e}}/\check{b}_{\rm{e}}$ in the coupling
condition on the right.

The coupled equilibrium agrees with what was found with the direct pressure
balance of Eq.\ (\ref{e:Patm2}) in Section \ref{s:fluxtube}.  The external
asymptotic longitudinal constituent function $\check{b}_{\rm{e}}$ must be
constant along flux tubes, like the constituent function $\check{b}$ itself by
the continuity relation (\ref{e:divb1}), and needs to be constant between flux
tubes too for the background field to be everywhere consistently defined. The
interconnectedness of flux tubes over the quiet corona thus suggests that
$\check{b}_{\rm{e}}$ must be a large-scale global constant, giving a constant
coupling coefficient $\beta_{\rm{c}}$ for their product to be constant, which is
consistent with the assumption from Section \ref{s:fluxtube} that flux tubes
must be similarly conditioned globally.  The model can accommodate solar active
regions with a gas pressure $P_{\rm{e}}$ that remains continuous crossing from
the quiet sun into the active region, with beta $\beta_{\rm{c}}$ very much
reduced and the background field ${\bf{B}}_{\rm{e}}$ correspondingly stronger.

As described in Section \ref{s:local} and in Appendix \ref{a:MHs1}, in
equilibrium solutions, with flows or temperature variations in embedded fields,
or in a non-potential background field, terms add to the Lorentz force, and so
to the curly braces in Eq.\ (\ref{e:MHs2c}) and to the local equilibrium in Eq.\
(\ref{e:MHs3a}).  However there is only one atmosphere, so an atmospheric
coupling relation like Eq.\ (\ref{e:MHs3b}), must have a unique global
functional form even with a mixture of embedded current-carrying fields with
their various and differing internal consistencies.  Added Lorentz-force terms
exhibit specific globally varying properties, but a most-significant-order
Lorentz force Eq.\ (\ref{e:Fi0b}) always appears with its single globally
varying part ${\bf\Omega}^2$ factored externally.  Thus whatever the mixture of
current-carrying fields, the gas-pressure coupling with ${\bf\Omega}^2$ in Eq.\
(\ref{e:MHs3b}) appears to be the only reasonable universally consistent
possibility, and so the coupled equilibrium of Eq.\ (\ref{e:MHs4b}).  Individual
embedded fields with differing internal contributions must adjust within to find
equilibrium balance along their lengths with any overall average-defined
background atmosphere.

\section{Local Equilibrium}
\label{s:local}

The relative gas pressure $\check{P}$ varies only locally to the embedded field,
defining its cross-sectional profile relative to the local ambient gas pressure.
It is constant across a force-free embedded field, but exhibits a fractional dip
in the center of a non-force-free embedded field of increased field strength, or
a fractional bump in one of decreased field strength.

The relative gas pressure $\check{P}$ must be conserved along the
current-carrying field lines for the local Eq.\ (\ref{e:MHs3a}) to be satisfied.
Its gradient is parallel to the transverse constituent function
${\bf\nabla}\check{P}\parallel{\bf\nabla}\check{a}$, and so must satisfy the
functional relation $\check{P}=\check{P}[\check{a}]$. With the continuity of
$\check{a}$ along the lines of the flux surface from Eq.\ (\ref{e:diva}), then
\begin{equation}
{\bf\Omega}{\bf\cdot}{\bf\nabla}\check{P}[\check{a}]=
\left({d\check{P}\over d\check{a}}\right){\bf\Omega}{\bf\cdot}{\bf\nabla}\check{a}=0,
{\rm\ or}\quad {\bf{B}}_{\rm{e}}{\bf\cdot}{\bf\nabla}\check{P}=0,
\label{e:divP}
\end{equation}
using the flux-surface background-field parallelism
${\bf\Omega}\parallel{\bf{B}}_{\rm{e}}$ in the rightmost equality.  The
constituent functions are normalized so that $\check{a}$ and $\check{b}$ are
actual magnetic field strengths and $\check{P}$ an actual gas pressure at a
reference height where $|{\bf\Omega}|^2=1$.  Away from that height, the
magnetic-field constituent functions $\check{a}$ and $\check{b}$ multiplied by
the flux surface $|{\bf\Omega}|$ give the actual transverse and longitudinal
field fluxes, and the relative gas pressure $\check{P}$ multiplied by
${\bf\Omega}^2$ the actual gas pressure.

With the functional constraint $\check{P}=\check{P}[\check{a}]$, the local
equilibrium relation Eq.\ (\ref{e:MHs3a}) takes the familiar form of the GS
equation
\begin{equation}
-{1\over4\pi}{\bf\nabla}^2\check{a}=
{d\over d\check{a}}\left(\check{P}+{1\over8\pi}\check{b}^2\right),
\label{e:GS0}
\end{equation}
but here written for 3D spatial functions.  The GS relation is essentially an
expression of the conservation of the total gas-plus-magnetic pressure in the
vicinity of the embedded field.  The constituent function can always be written
$\check{a}=\check{a}[\xi]$ with an intermediate variable $\xi$, which is a
function of the embedded-field coordinates, e.g.\ Cartesian $\xi=\xi[x,y,z]$ or
cylindrical $\xi=\xi[r,\phi,z]$.  With the intermediate functional dependence,
the Laplacian is expanded
\begin{equation}
{\bf\nabla}^2(\check{a}[\xi])={d^2\check{a}\over d\xi^2}|{\bf\nabla}\xi|^2+{d\check{a}\over d\xi}{\bf\nabla}^2\xi,
\label{e:remap0}
\end{equation}
and with $\check{b}=\check{b}[\check{a}[\xi]]=\check{b}[\xi]$, and
$\check{P}=\check{P}[\check{a}[\xi]]=\check{P}[\xi]$, the GS Eq.\ (\ref{e:GS0})
becomes
\begin{equation} 
{d\over d\xi}\left(\check{P}[\xi]+{1\over8\pi}\check{b}[\xi]^2\right)
+{1\over8\pi}{d\over d\xi}\left(d\check{a}\over d\xi\right)^2|{\bf\nabla}\xi|^2
+{1\over4\pi}\left({d\check{a}\over d\xi}\right)^2{\bf\nabla}^2\xi=0,
\label{e:GS1}
\end{equation}
applying the chain rule.  Integrating in $\xi$ between any $\xi_0$ and $\xi_1$
yields
\begin{equation} 
\check{P}[\xi]+{1\over8\pi}\check{b}[\xi]^2
+{1\over8\pi}\left({d\check{a}\over d\xi}\right)^2
+{1\over4\pi}\int_{\xi_0}^{\xi_1}{\left({d\check{a}\over d\xi}\right)^2{\bf\nabla}^2\xi d\xi}
={\rm{constant}},
\label{e:GS2}
\end{equation}
taking $|{\bf\nabla}\xi|^2=1$, which is relevant for Cartesian current-sheet and
cylindrical axisymmetric (not azimuthally varying) flux-tube solutions.  The
remaining integral can be seen to represent an added pressure due to magnetic
field tension in the current-carrying field.

In local embedded-field equilibria of interest, internal magnetic field tension
is negligible.  For current-sheets, $|{\bf\nabla}\xi|^2=1$ and
${\bf\nabla}^2\xi=0$, and the added integral term in Eq.\ (\ref{e:GS2})
disappears.  The condition ${\bf\nabla}^2\xi\rightarrow0$ is asymptotic for
thin-sheathed flux tubes.  Multiplying through by the flux surface
${\bf\Omega}^2$ renormalizes the constant constituent functions along their
paths through the corona, giving a conserved total pressure
\begin{equation}
P+{{\bf{B}}^2\over8\pi}={\rm{constant}},
\label{e:GS3}
\end{equation}
around every spatial locale, substituting using $\check{P}=P/{\bf\Omega}^2$ from
Eqs.\ (\ref{e:Pidefa}) and (\ref{e:MHs3b}), with
${\bf{B}}^2=((d\check{a}/d\xi)^2+\check{b}^2){\bf\Omega}^2$ from Eq.\
(\ref{e:B0}), and ignoring the relatively small derivatives of the flux surface
${\bf\Omega}$.  Such a conserved total pressure was assumed in Eq.\
(\ref{e:Pdif}) for direct pressure balance in the vicinity of current-carrying
fields in Section \ref{s:fluxtube}. A significant pressure offset from a direct
pressure balance within the current-carrying field is possible due to the added
tension term in Eq.\ (\ref{e:GS2}), which may be relevant for flux tubes with
inhomogeneous internal structure as below the coronal base, as described further
below.

Near force-free current-carrying fields follow the same relations but with
relatively small gas-pressure changes compared to magnetic-pressure values,
i.e.\ $|P_{\rm{i}}-P_{\rm{e}}|\ll{\bf{B}}_{{\rm{e}}}^2/(8\pi)$ and
${\bf{B}}_{\rm{i}}^2/(8\pi)$. Pure force-free fields exhibit current-density
magnetic-field parallelism ${\bf{J}}\parallel{\bf{B}}$ or
${\bf\nabla}{\bf\times}{\bf{B}}=\alpha{\bf{B}}$ for $\alpha$ constant along
field lines. Substituting ${\bf{B}}$ from Eq.\ (\ref{e:B0}) and
${\bf\nabla}{\bf\times}{\bf{B}}$ from Eq.\ (\ref{e:J0}) with ${\bf\Omega}$ a
constant vector shows that $\alpha=d\check{b}/d\check{a}$.  In the force-free
limit, the derivative vanishes $d\check{P}/d\check{a}=0$ in the GS Eq.\
(\ref{e:GS0}) meaning both $P$ and ${\bf{B}}^2$ are separately constant in every
locale in Eq.\ (\ref{e:GS3}), results consistent with what is expected for a
vanishing Lorentz force in force-free fields.

GS equilibria are explored by looking for functions $\check{a}=\check{a}[\xi]$
that satisfy the functional constraint
${\bf\nabla}^2\check{a}=({\bf\nabla}^2\check{a})[\check{a}]$. The functional
constraint $\check{b}=\check{b}[\check{a}]$ limits the choices for $\check{b}$
for a given spatial function $\check{a}$, but the constraint
${\bf\nabla}^2\check{a}=({\bf\nabla}^2\check{a})[\check{a}]$ is a template for
the GS Eq.\ (\ref{e:GS0}) itself and defines the allowed functional forms for
$\check{a}$.  Given a spatial dependence $\check{a}=\check{a}[\xi]$, as long as
${\bf\nabla}^2\check{a}=({\bf\nabla}^2\check{a})[\xi]$, it can be seen that
${\bf\nabla}^2\check{a}=({\bf\nabla}^2\check{a})[\check{a}]$.  The derivatives
$d^2\check{a}/d\xi^2$ and $d\check{a}/d\xi$ in Eq.\ (\ref{e:remap0}) are always
functions of $\xi$ alone, so there is always complete freedom in choosing the
outer function $\check{a}[\xi]$.  However with an arbitrary coordinate
dependence in $\xi$, the derivative factors $|{\bf\nabla}\xi|^2$ and
${\bf\nabla}^2\xi$ need not be functions of $\xi$ alone.

Many useful continuous functions for $\check{a}[\xi]$ come to mind, like some
which have an explicit inverse $\xi=\xi[\check{a}]$, as a Gaussian
$\check{a}[\xi]=\check{a}_{\rm{i}}\exp[-\xi^2]$ for $\check{a}_{\rm{i}}$ its
central internal $\xi=0$ value, with its double-valued inverse
$\xi[\check{a}]=\pm(\ln[\check{a}/\check{a}_{\rm{i}}])^{1/2}$, or the smooth
cutoff $\check{a}[\xi]=\check{a}_{\rm{i}}/(1+\exp[\xi])$ with its single-valued
inverse $\xi[\check{a}]=\ln[\check{a}/\check{a}_{\rm{i}}-1]$.  However the
inverse need not be explicit to give an allowed outer function $\check{a}[\xi]$,
as with a sinc function $\check{a}[\xi]=\sin[\xi]/\xi$, with its multivalued
transcendental inverse $\xi[\check{a}]$.

Two of the 13 geometries separable under the Laplacian operator
(\citealt{Moon+Spencer1971}, \S1) are mainly of relevance for embedded
current-carrying fields:\ the Cartesian $(x,y,z)$ current sheet in a background
field with 1D symmetry as in a linear arcade, and the cylindrical $(r,\phi,z)$
flux tube in a background field with 2D symmetry around the flux-tube center.
Other forms such as spherical arise too as when considering embedded plasmoids.

In straight field geometries with ${\bf\Omega}\rightarrow z$, for variations
perpendicular to $z$ across a Cartesian current sheet $\xi\rightarrow x$, the
derivative factors in Eq.\ (\ref{e:remap0}) are $|{\bf\nabla}\xi|^2=1$ and
${\bf\nabla}^2\xi=0$.  For radial variations in a cylindrical flux tube constant
in $z$, $\xi\rightarrow r$, corresponding to a twisted magnetic field, the
derivative factors become $|{\bf\nabla}\xi|^2=1$ and
${\bf\nabla}^2\xi=1/r=1/\xi$.  In both cases
$|{\bf\nabla}\xi|^2=(|{\bf\nabla}\xi|^2)[\xi]$ and
${\bf\nabla}^2\xi=({\bf\nabla}^2\xi)[\xi]$, so these geometries always satisfy
the functional constraint
${\bf\nabla}^2\check{a}=({\bf\nabla}^2\check{a})[\check{a}]$ and the GS Eq.\
(\ref{e:GS0}) with any continuous outer function $\check{a}[\xi]$.

A nonaxisymmetric flux tube with purely azimuthal variations
$\xi\rightarrow\phi$ in a straight field geometry exhibits derivatives
$|{\bf\nabla}\xi|^2=1/r^2$ and ${\bf\nabla}^2\xi=0$, giving
${\bf\nabla}^2(\check{a}[\xi])$ a residual $r$ dependence, which precludes the
functional dependence $|{\bf\nabla}\xi|^2=(|{\bf\nabla}\xi|^2)[\xi]$, and so the
isotropy constraint ${\bf\nabla}^2\check{a}=({\bf\nabla}^2\check{a})[\check{a}]$
is not satisfied. However straight-field nonaxisymmetric solutions can be
built-up based upon the complex $\xi=x+iy$ that do satisfy the functional
dependencies giving
${\bf\nabla}^2\check{a}=({\bf\nabla}^2\check{a})[\check{a}]$. These lead to
azimuthally tessellated solutions, which, if confined to the outer sheath,
resemble sunspot penumbrae.  An exploration of these solutions is outside the
scope of this paper.

Projecting the cross-sectional 2D profile in $\check{a}$ and $\check{b}$ in the
embedded field from a given reference height along the lines of the unperturbed
background field into 3D, consistent with the continuity requirements Eqs.\
(\ref{e:diva}) and (\ref{e:divb1}), strongly constrains the functional
$({\bf\nabla}^2\check{a})[\check{a}]$.  Most 3D forms cannot represent viable
solution forms because they give slowly changing spatial derivatives in the
Laplacian ${\bf\nabla}^2\check{a}$ as the embedded field varies in
cross-sectional area or shape, while $\check{a}$ remains constant along the
lines of the background field.

\begin{figure}[ht!]
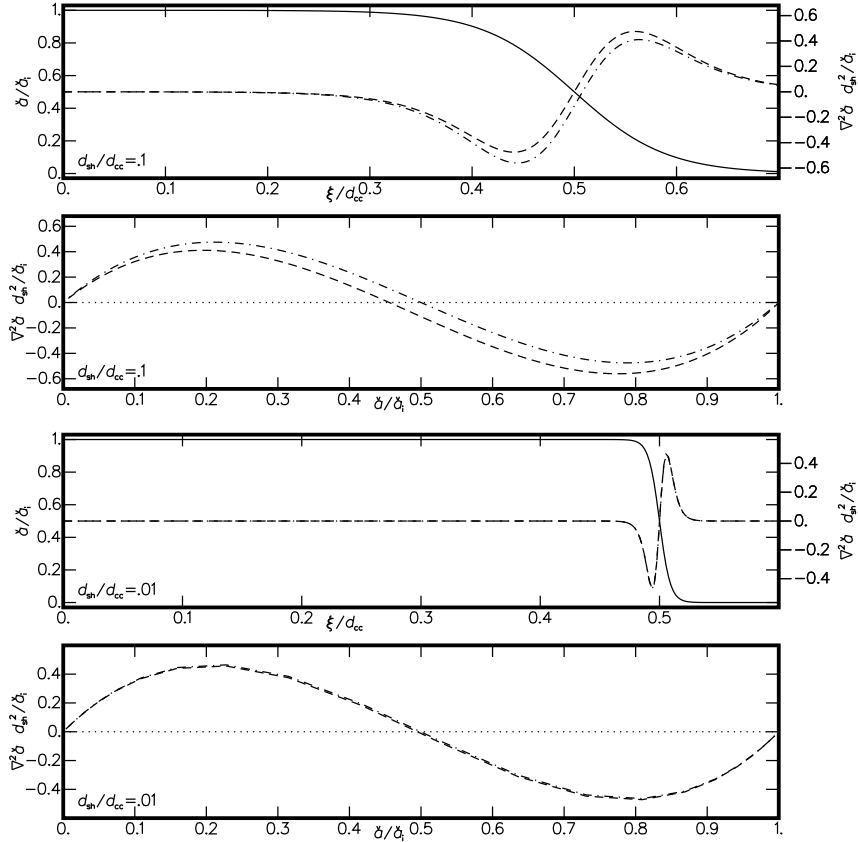

\centering\noindent
\centerline{\includegraphics[width=0.7\columnwidth,bb=28 277 612 564]{ftprof100.eps}} 
\centerline{\includegraphics[width=0.7\columnwidth,bb=28 277 612 564]{ftprof010.eps}} 
\caption{Profile $\check{a}[\xi]$ (solid) with $({\bf\nabla}^2\check{a})[\xi]$
for a flux tube with $\xi=r$ (dashed lines) and a current sheet with $\xi=\pm x$
(dot-dashed lines) (labeled on right), with the corresponding functional
$({\bf\nabla}^2\check{a})[\check{a}]$ (second and fourth panels), for sheaths of
relative thicknesses $d_{\rm{sh}}/d_{\rm{cc}}$.  The functional can preserve its
form even as the sheath thins, with the flux-tube and current-sheet profiles
$({\bf\nabla}^2\check{a})[\xi]$ and functionals
$({\bf\nabla}^2\check{a})[\check{a}]$ becoming asymptotically alike.}
\label{f:ftprof}
\end{figure}

A pliable 3D solution form in a smoothly changing embedded field is described
with a Laplacian ${\bf\nabla}^2\check{a}$ that is a function of the local value
of $\check{a}$, even as the underlying spatial function $\xi$ in Eq.\
(\ref{e:remap0}) becomes less well-defined.  Variations in the constituent
function $\check{a}$ over the 2D cross section of the embedded field might be
confined to a thin, but finite, outer sheath that maintains constant thickness in
the embedding field even as it spreads out or changes shape.  With the function
$\check{a}[\xi]$ varying in 1D $\xi$, the distance locally perpendicular to the
plane of the sheath, the Laplacian $({\bf\nabla}^2\check{a})[\xi]$ from Eq.\
(\ref{e:remap0}) passes through its full range of values in the thin sheath as
illustrated by hypothetical 1D cross sections in Figure \ref{f:ftprof}.  The
shape of the functional $({\bf\nabla}^2\check{a})[\check{a}]$ is determined by
the detailed cross section of $\check{a}$ through the thin sheath, with its
amplitude scaling inversely with the sheath width as $1/d_{\rm{sh}}^2$.  For a
given spatial function $\check{a}[\xi]$, the Laplacian
$({\bf\nabla}^2\check{a})[\check{a}]$ in a cylindrical flux tube differs only
slightly from that of a current sheet by the small order
$d_{\rm{sh}}/d_{\rm{cc}}$, as illustrated by the convergence of the profiles
from the upper to the lower pair of panels in the figure.

With a constant sheath width $d_{\rm{sh}}$, the Laplacian
$({\bf\nabla}^2\check{a})[\check{a}]$ can remain consistently defined all along
the length of the embedded field.  With all of the variation in $\check{a}$
confined to the thin sheath, any overall reshaping of the embedded-field cross
section along the background field lines away from its nominal geometric
equilibrium, or curvature effects as field lines spread, might be corrected with
small deviations in the trajectory of the flux surface ${\bf\Omega}$ from the
unperturbed background potential field lines within the sheath.  Field-line
deviations between the flux surface and the background field may cause
background-field perturbations within the embedded field.  Background field
deviations, or any remaining misalignment between the flux surface and
background magnetic field, both add pressure anisotropies to the GS Eq.\
(\ref{e:GS0}) of the same relative order $d_{\rm{sh}}/H_{\rm{Be}}$.  Added
anisotropic terms come from the Lorentz-force term $F_{\rm{e}}$ in Eq.\
(\ref{e:Fe1}) and from $F_{\rm{i1}}$ in Eq.\ (\ref{e:Fi1a}) as discussed in
Appendix \ref{a:Fi}.

As all of the $\check{a}$ variation is in the thin sheath, so is all of the
transverse magnetic field and the flux-tube field twist, the cylindrical
${\bf{B}}_{\phi}$ component in the definition Eq.\ (\ref{e:B0}).  The minimum
sheath width possible is determined by the relative central transverse
$\check{a}_{\rm{i}}$ to longitudinal $\check{b}_{\rm{i}}$.  The introduced terms
may cause unsteady behavior, flows, or small temperature and density variations
within the sheath.

The constituent functions $\check{a}$ and $\check{b}$ are always strictly
conserved along the length of the embedded field, however small changes in the
gas pressure $\check{P}$ in the sheath itself may be incurred.  Anisotropic
Lorentz forces may change the pressure difference across the sheath, which add
integral terms like the tension term in Eq.\ (\ref{e:GS2}).  In the interior of
the embedded field, anisotropic effects are vanishing, strictly in a Cartesian
current sheet in a potential background field, and to high order otherwise, as
the derivatives of the constituent function $\check{a}$ are negligible there. In
static thermally uniform conditions, relative pressure differences within the
sheath can be seen to go like the Lorentz-force deviations
$d_{\rm{sh}}/H_{\rm{Be}}$.  In solar flux tubes of 0.7 Mm diameter with a 10\%
sheath, the relative pressure fluctuations are of the small relative order
$d_{\rm{sh}}/H_{\rm{Be}}\simeq4$E$-4$.

In a current-carrying field embedded in a background {\it non}-potential
magnetic field, the Lorentz force in Eq.\ (\ref{e:Fi0b}) becomes augmented with
additional terms not necessarily parallel to ${\bf\nabla}\check{a}$ giving
anisotropic pressure effects of order $d_{\rm{cc}}/H_{\rm{Be}}$ over the cross
section of the current-carrying field, as elaborated in Appendix \ref{a:MHs1}.
With flux-tube flows, or ambient temperature and density differences, terms of
larger order than $d_{\rm{cc}}/H_{\rm{Be}}$ may be introduced.  Though the
transverse and longitudinal constituent functions $\check{a}$ and $\check{b}$
are always conserved along field lines, variations in the relative gas pressure
$\check{P}$ may be introduced of order the anisotropies.  Solar prominences
appear to be thicker embedded coronal flux tubes with
$d_{\rm{cc}}/H_{\rm{Be}}\simeq1/20$, exhibiting high-speed counter flows in
their outer sheaths \citep{LinY+Engvold+Wiik2003}, and very much higher internal
densities and cooler temperatures than the local ambient.

Even in the absence of an embedding background magnetic field, as below the
solar coronal base, current-carrying fields will exhibit similarly conserved
properties.  As long as a flux tube is adequately represented by Eq.\
(\ref{e:B0}) with a conserved flux surface ${\bf\Omega}$, a property of the
mathematical decomposition, the constituent functions $\check{a}$ and
$\check{b}$ are conserved along the flux surface ${\bf\Omega}$.  The
separability argument of Section \ref{s:MHs} and Appendix \ref{a:MHs1} follows,
which leads to an ambient gas pressure related to the flux surface
$P=\check{P}{\bf\Omega}^2$ from Eqs.\ (\ref{e:Pidefa}) and (\ref{e:MHs3b}), and
then the equilibrium $P_{\rm{e}}\sim{\bf\Omega}^2$, without refering to any
background field ${\bf{B}}_{\rm{e}}$.  Though pressures are still coupled, where
conditions are not magnetically dominated, the ambient gas pressure controls the
properties of the current-carrying field, {\it not} the current-carrying field
the ambient gas pressure. With variations in the cross section of the
current-carrying field along its length, sheathed solutions are still expected.

Where flux tubes are subjected to localized external forcings as due to opposing
high-density material flows, internal counter Lorentz forces must be produced,
and magnetic flux may be being created or destroyed.  Solar photospheric flux
tubes are located in opposing granulation flows, and are known to exhibit very
inhomogeneous cross sections \citep{Keller1995, BergerTE+++1998b}, suggesting
large distortions from a normal GS equilibrium.  With the spreading of a
photospheric flux tube in the falling solar atmospheric gas pressure above,
spatial derivatives in the internal magnetic field ${\bf{B}}$ are reduced and so
correspondingly the current density
${\bf{J}}=\sfrac{c}{4\pi}{\bf\nabla}{\bf\times}{\bf{B}}$ and Lorentz force,
which is consistent with what is seen in solar chromospheric magnetograms
\citep{Metcalf+++1995}.  With a uniform spreading, the relative gas pressure
though should still be reasonably conserved with the flux-tube gas-pressure
contrast varying in proportion to the ambient.  However with internal blending
and mixing with adjacent flux tubes, it too may be significantly reduced.

Rather than spreading and adding to the large-scale background field uniformly
with the cancellation of all its internal nonpotential magnetic field, some flux
tubes, or at least internal knots of twisted field within them, may reach the
solar or stellar coronal atmosphere above largely intact to produce embedded
non-force-free flux tubes.  \cite{LinY+Engvold+++2005} do identify individual
flux tubes as small as 200 km in diameter in the inter-granular lanes in the
solar photosphere, traceable intact through the chromosphere in H$\alpha$, and
into the corona above, where they smoothly join with much larger-diameter
prominence flux tubes. As non-force-free flux tubes embedded in a near-potential
magnetic field are a viable equilibrium arrangement, their presence must be
considered a reasonable possibility within the wider scope of astrophysical
processes in planets, neutron stars, black holes, galaxies, galactic
clusters, or the larger cosmological system.

\section{Global Equilibrium}
\label{s:glob}

The global pressure balance obtained in Section \ref{s:MHs} and Appendix
\ref{a:MHs1} is the usual MHD equation as if no embedded fields were present,
except that the gas pressure is everywhere constrained to vary with the
background magnetic field.  It is rewritten from Eqs.\ (\ref{e:MHs4b}) or
(\ref{e:MHs4x}) as
\begin{equation}
{1\over H_{\rm{s}}g[R]}{D{\bf{u}}\over Dt}
=-{\bf\nabla}\ln[{\bf{B}}_{\rm{e}}^2]-{1\over H_{\rm{s}}}{\bf\nabla}R,
\label{e:flow1}
\end{equation}
using the coupled ambient magnetic pressure substituted for the ambient gas
pressure.

In flowless magnetically coupled conditions, the global equilibrium contains the
hydrostatic radial balance of pressure and gravity, which leads to specific
radial density and temperature profiles. The hydrostatic relation, the radial
MHD pressure Eq.\ (\ref{e:flow1}) without flows, is written
\begin{equation}
-{\partial\ln[{\bf{B}}_{\rm{e}}^2]\over\partial R}
={1\over H_{\rm{s}}}={\bar{m}g[R]\over kT_{\rm{e}}},
\label{e:corHT2}
\end{equation}
substituting the definition for the ambient static pressure scale height
$H_{\rm{s}}=P_{\rm{e}}/(\rho_{\rm{e}}g[R])$ with the perfect-gas law
$P_{\rm{e}}=\rho_{\rm{e}}kT_{\rm{e}}/\bar{m}$ across the rightmost equality,
with $\rho_{\rm{e}}$ the ambient coronal density; thus
\begin{equation}
kT_{\rm{e}}=\sfrac{1}{2}\bar{m}g[R] H_{\rm{Be}},
\label{e:corHT3}
\end{equation}
where $H_{\rm{Be}}=-(\partial\ln[|{\bf{B}}|]/\partial R)^{-1}$.

With a single-degree spherical harmonic $\ell$, the magnetic field exhibits a
separable radial dependence that goes as $1/R^{\ell+2}$ in all of its vector
elements
${\bf{B}}_{\rm{e}}={\bf{B}}_{{\rm{s}}\ell}[\Theta,\Phi](R_{\rm{s}}/R)^{\ell+2}$,
with $R_{\rm{s}}$ the object radius, and the vector surface field
${\bf{B}}_{{\rm{s}}\ell}[\Theta,\Phi]$ in colatitude $\Theta$ and longitude
$\Phi$ over the globe, which is a sum over the spherical-harmonic index $m$ in
spherical-harmonic expansions (e.g.\ \citealt{Altschuler+Newkirk1969}, Eqs.\
(8)-(10)).  Substituting the single-degree spherical-harmonic magnetic field
into Eq.\ (\ref{e:corHT2}) gives the scale height $H_{\rm{Be}}=R/(\ell+2)$ and
so the temperature
\begin{equation}
kT_{\rm{e}}={\bar{m}g_{\rm{s}}R_{\rm{s}}^2\over(2\ell+4)R}=
{GM_{\rm{s}}\bar{m}\over(2\ell+4)R},
\label{e:corHT4}
\end{equation}
using the inward gravitational acceleration at the photospheric surface
$g[R]=g_{\rm{s}}(R_{\rm{s}}/R)^2$, and substituting with the surface gravity
$g_{\rm{s}}=GM_{\rm{s}}/R_{\rm{s}}^2$ across the second equality, for $G$ the
universal gravitational constant and $M_{\rm{s}}$ the total object interior
mass.

With a solar coronal temperature of $T_{\rm{e}\sun}=1.6$E6K near the solar
photosphere $R=R_{\rm{s}}=R_\sun$, the equation gives a solar spherical harmonic
of degree $\ell_\sun=2.41$, with $g_{\rm{s}}=g_\sun$ or solar mass
$M_{\rm{s}}=M_\sun$, and mean particle mass $\bar{m}=0.602m_{\rm{p}}$ for solar
coronal ionization abundances with proton mass $m_{\rm{p}}$. Static coronal
temperatures in a potential field of single spherical harmonic drop off like
$1/R$. The same solar coronal temperature is found at a height in the corona
$R=1.2R_{\sun}$ with a spherical-harmonic degree $\ell=1.68$, or at height
$R=1.3R_{\sun}$ with $\ell=1.4$.  The solar coronal magnetic field is known to
feel its most significant contributions from low-degree near dipolar spherical
harmonics $\ell\gtrsim1$ \citep{Altschuler+++1974}.

With a superposition of spherical harmonics in degree $\ell$, temperature is
near constant in static solutions out to about $R=2R_{\rm{s}}$, and then slowly
falls off asymptotically above, following the profile of the smallest
significant spherical-harmonic degree $\ell$ in the superposition.  It drops to
about half its surface value at $R=4R_{\rm{s}}$ for a dominantly dipolar
magnetic field, as elaborated in other work to be submitted.

For a given superposition of spherical harmonics, coronal temperature profiles
scale simply with object mass over radius
\begin{equation}
T_{\rm{e}}={(M_{\rm{s}}/M_{\sun})\over(R_{\rm{s}}/R_{\sun})}T_{\rm{e}\sun}.
\label{e:corHT5}
\end{equation}
Scaling in mass over radius from a given average solar coronal temperature
$T_{\rm{e}\sun}$ gives an average coronal temperature for magnetic objects
supposing a similar mean particle mass $\bar{m}$ and background-field
spherical-harmonic distribution, mainly dipolar but with power extending to high
spherical harmonics.  Unresolved coronal temperatures must be most influenced by
temperatures within a few radii above the coronal base where densities are still
high.  Although the sun's base coronal temperature during quiet conditions is
about 1.6E6K, its average coronal temperature is somewhat lower, taken here to
be 1E6K consistent with unresolved quiet solar coronal X-ray inferences
(\citealt{Orlando+Peres+Reale2004}, Figure 6; \citealt{Argiroffi+++2008}, Figure
4).

Main-sequence stars exhibit a slowly increasing ratio of mass to radius in the
Hertzsprung-Russell (HR) diagram from late M type to early O type, with their
equilibrium coronal temperatures ranging correspondingly from 1E5K to near 1E7K,
extrapolating from a mean solar coronal temperature $T_{\rm{e}\sun}=1$E6K in
Eq.\ (\ref{e:corHT5}). In moving to higher luminosities and cooler surface
temperatures from the main sequence up through the giant and supergiant branches
from the sun, stellar mass scales up some but radius much more.  The giant
$\alpha$ UMa (K0 II-III) sits on the \citep{Linsky+Haisch1979} coronal dividing
line.  Its mass is $M_{\rm{s}}=3.70M_{\sun}$ and radius
$R_{\rm{s}}=29.6R_{\sun}$ as a visual binary (\citealt{GrayDF2005}, Table 15.1),
or consistently following evolutionary tracks (e.g.\
\citealt{Maeder+Meynet1989}, Figure 15), which gives an equilibrium coronal
temperature $T_{\rm{e}}=1.25$E5K.  A temperature of 1E5K is consistent with the
spectral diagnostics used to define the dividing line.

With higher luminosity and cooler surface temperature in the supergiant branches
beyond the coronal dividing line, stars exhibit yet cooler predicted coronal
temperatures, as the representative $\alpha$ Ori (M2 Iab) with an estimated mass
$M_{\rm{s}}=19M_{\sun}$ and radius $R_{\rm{s}}=800R_{\sun}$
(\citealt{LangKR1992}, Section 9.6), which gives a coronal temperature
$T_{\rm{e}}=24000$K.  Such a cool corona might be called an `extended
chromosphere', like what is inferred from visible spectra of cool giants and
supergiants \citep{Weymann1962}, from high-resolution interferometric
observations (\citealt{Hebden+Eckart+Hege1987}, and references), and from
space-based UV observations \citep{Carpenter+BrownA+Stencel1985}.  A more
thorough elaboration of stellar comparisons is being prepared for separate
publication.

Mass and radius scale upward approximately similarly to very much larger
galactic dimensions, to still give average equilibrium coronal temperatures
somewhat less than central main-sequence or solar values.  Taking the lower
boundary for the galactic potential magnetic field of the Milky Way to be at the
upper limit of the bulge solid-body rotation curve at $R_{\rm{s}}=1.8$
kpc$=8$E$10\ R_{\sun}$, where an accepted estimate for the contained (baryonic)
mass is $M_{\rm{s}}=1.5$E$10\ M_{\sun}$ (e.g.\ \citealt{Strobel2007}, Section
14.2.4), gives an average equilibrium coronal temperature of
$T_{\rm{e}}=2$E$5$K.  At a solar galactic radial distance of about $R=8$
kpc$=4.4R_{\rm{s}}$, the local estimated coronal temperature is 1.6E5K, a little
lower than the average, scaling from a solar surface coronal temperature of
1.6E6K.  A slightly higher temperature of $2-3$E5K is inferred from
measurements, but with inhomogeneities, consistent with some additional heating
by supernovae remnants \citep{JenkinsEB+Meloy1974, JenkinsEB1978b,
Savage+++2003}. A significant cosmic-ray pressure should not alter galactic
temperature estimates as discussed at the end of Appendix \ref{a:MHs1}, and
differing galactic structural elements should not change its gaseous
thermodynamics within the MHD approximation.

Planetary coronal estimates are complicated by their varying atmospheric
chemical compositions, and by the differing natures of their magnetospheres.
Their cooler atmospheres lead to an increased and spatially varying mean
particle mass $\bar{m}$ and increased temperature estimate in Eq.\
(\ref{e:corHT3}).  Some planetary bodies have negligible internal magnetic
fields, but still exhibit induced magnetospheres in the solar interplanetary
field and wind.  On the near-sun day side, planetary magnetospheres are
compressed and strengthened, which may decrease the magnetic-field scale height
$H_{\rm{Be}}$ and temperature estimate, while increasing the density estimate.
Whereas on the far-sun night side magnetospheres are stretched out and weakened,
which increases $H_{\rm{Be}}$ and correspondingly the temperature estimate,
while decreasing the density estimate. Also the effectiveness of non-force-free
flux tubes in providing energy for planetary atmospheres may be limited because
their sources of energy may be distantly removed, as due to footpoint motions in
the solar interior.  Elaboration of these effects is needed before predictions
of planetary magnetostatic atmospheres can be adequately presented, and so that
development must be postponed here.

Neutron stars have very large ratios of $M_{\rm{s}}/R_{\rm{s}}$.  With typical
values $M_{\rm{s}}=1.4M_{\sun}$ and $R_{\rm{s}}=10$ km, a temperature of
$T_{\rm{e}}=1$E11K $=8400$ keV is expected for a possible corona.  It is
believed that rotating black holes may develop strong surrounding potential
magnetic fields \citep{Karas+++2013}.  If embedded flux tubes are present, with
the Schwarzschild radius $R_{\rm{s}}=2GM_{\rm{s}}/c^2$ for their contained mass
$M_{\rm{s}}$, any size black hole would exhibit the coronal temperature
\begin{equation}
T_{\rm{e}}={c^2\over2G}{R_{\sun}\over M_{\sun}}T_{\rm{e}\sun}=
2.4{\rm{E}}11{\rm{K}}=2{\rm{E}}4\ {\rm{keV}},
\label{e:corHT6}
\end{equation}
about twice that of a typical neutron star.

\begin{figure}[ht!]
\centering\noindent
\centerline{\includegraphics[width=0.7\columnwidth]{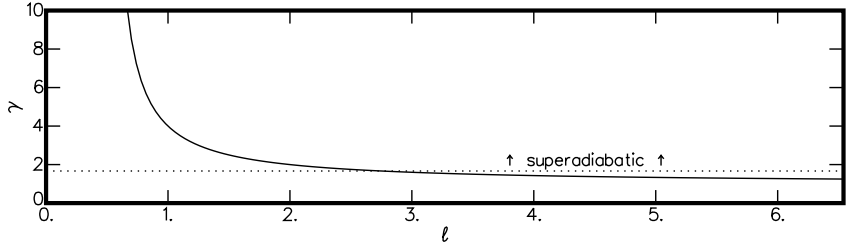}} 
\caption{Polytrope index $\gamma$ verses spherical-harmonic degree $\ell$ for
the coronal atmosphere with the adiabatic line $\gamma=5/3$ shown.}
\label{f:poly}
\end{figure}

The static equilibrium of a potential magnetic field of single-degree spherical
harmonic can be seen to be a polytrope. The gas pressure satisfies
$P_{\rm{e}}[R]\sim{\bf{B}}_{\rm{e}}^2\sim1/R^{2\ell+2}$, and the density
$\rho_{\rm{e}}/\bar{m}=P_{\rm{e}}/(kT_{\rm{e}})\sim1/R^{2\ell-1}$ using the
perfect-gas law. Combining yields the polytrope relation
$P_{\rm{e}}\sim\rho_{\rm{e}}^\gamma$ with
\begin{equation}
\gamma={2\ell+2\over2\ell-1}.
\label{e:poly1}
\end{equation}
The polytrope index $\gamma$ is plotted as a function of spherical-harmonic
degree $\ell$ in Figure \ref{f:poly} with $\gamma\rightarrow\infty$ as
$\ell\rightarrow0.5$, $\gamma=4$ for $\ell=1$, and $\gamma\rightarrow1$ as
$\ell\rightarrow\infty$.  The atmosphere is superadiabatically stratified with
$\gamma>5/3$, which occurs for degree $\ell<11/4$ as shown in the figure.
Though a superadiabatic gas alone would be convective, its coupling to the
large-scale potential magnetic field should be a stabilizing influence.

Coronal dissipation as cooling and material downflow give a deviation from
equilibrium as a reduced temperature or static scale height compared to the
magnetic-field scale height
$H_{\rm{s}}<H_{\rm{Be}}/2=-1/(\partial\ln[B_{\rm{e}}^2]/\partial R)$ in the
second term on the right side of Eq.\ (\ref{e:flow1}).  The right side of the
equation in its radial vector component $2/H_{\rm{Be}}-1/H_{\rm{s}}$ thus
becomes negative, corresponding to a decelerating positive radial outflow and
coronal matter deposition.  The outflow abates as the temperature and scale
height $H_{\rm{s}}=kT/(\bar{m}g)$ increase and the equilibrium is restored.
Without large turbulent effects, the main contribution to the heating in the
energy equation must come from gas compression.

Steady accelerated outflows are similarly necessitated over weak-field
low-pressure regions where ${\bf{B}}_{\rm{e}}^2$ is small over the globe, like
the accelerating wind outflow seen in solar coronal holes. The radial
gravitational pressure, the rightmost term, is zero in the horizontal component
of Eq.\ (\ref{e:flow1}).  The magnetic pressure in static regions of stronger
magnetic field strength outside the coronal hole must thus balance the Bernoulli
pressure for steady flows, which comes from the irrotational flow component
${\bf\nabla}{\bf{u}}^2/2$ in $D{\bf{u}}/Dt$.

At some height, perhaps not even much above the photospheric surface, the
accelerating wind outflow will become dominant over the magnetic field, which
may shred it and segregate it from the flow, consistent with a lack of internal
flow, as is seen in isolated relatively high-density plumes in solar coronal
holes \citep{Hassler+++1997}.  Though mixed flow and field regions are possible,
parallel frozen-in flow and field trajectories, as required by the steady
magnetic-induction equation in Eq.\ (\ref{e:ucont}), will mostly be
incommensurate in any mixed steady-flow and magnetic-field solution. Though
magnetic fields may be segregated from the high-speed flow, with the gas
pressure and magnetic field no longer coupled, the outflow remains still needed
for steady horizontal pressure balance with the high gas pressure in the
adjacent magnetically coupled flowless atmosphere.  Additional outward pressure
as due to radiation may of course alter the strength and spatial distribution of
the wind outflow.

\section{Conclusions}
\label{s:conc}

A large-scale spatially coherent magnetic field embedded with isolated
current-carrying fields is shown to be an MHD equilibrium configuration.
Current-carrying channels must run parallel to the embedding background magnetic
field overall, for a minimum Lorentz-force anisotropy consistent with the MHD
approximation. Then by direct pressure balance, a minimum atmosphere is
necessitated, described by the coupled external ambient gas and magnetic
pressures
\begin{equation}
P_{\rm{e}}=\beta_{\rm{c}}{{\bf{B}}_{\rm{e}}^2\over8\pi},
\label{e:PB}
\end{equation}
for $\beta_{\rm{c}}$ a single global constant corresponding to the relative
magnetic field strength of the overall strongest globally distributed embedded
non-force-free current-carrying fields.  Solar active-region fields are very
much stronger than quiet-sun fields, but their sources of energy due to
underlying footpoint motions must be much more limited because of their small
global areal coverage.  They should thus tend to relax to near force-free to be
in gas-pressure balance with the surrounding quiet global coronal atmosphere.

The MHD equilibrium separates into global and local pressure relations.  The
local embedded-field relation, in the ideal limit of an isotropic Lorentz force,
is described by the GS equation usual for straight 2D geometries, but with 3D
constituent functions representing the relative embedded transverse and
longitudinal field strengths, and the gas-pressure contrast compared to the
ambient. The GS relation is a functional equation, and its functional property
determines its solution form in 2D or 3D. Embedded-field 3D solutions are found
with the transverse field confined to a thin but finite outer sheath of constant
thickness, which contains all of the field twist in cylindrical flux tubes.  Due
to magnetic field continuity and gauge properties, the magnetic constituent
functions are always strictly conserved along the length of the embedded field,
corresponding to conserved transverse and longitudinal magnetic fluxes.  The
relative gas-pressure contrast is determined by the GS relation and stays mostly
uniform along the embedded field, however may show deviations due to
Lorentz-force anisotropies necessitated within the sheath; these are within the
order of the relative sheath thickness: $d_{\rm{sh}}/H_{\rm{Be}}\simeq4$E$-4$
for the finest observed lower-solar-coronal flux tubes of diameter
$d_{\rm{cc}}=0.7$ Mm in a sheath of relative thickness
$d_{\rm{sh}}/d_{\rm{cc}}\simeq10$\%.

The global relation retains large-scale temporally changing and steady flows,
temperature variations, gravity, and Lorentz forces due to the global background
magnetic field, but loses all terms varying locally in embedded current-carrying
fields.  Instead the equilibrium is constrained by the same coupling between the
ambient gas pressure and large-scale background field expressed in Eq.\
(\ref{e:PB}) from the direct pressure-balance argument.  The global relation is
independent of the specific properties of the embedded fields, their geometric
character as current-sheet or flux tube, their internal flows, or ambient
density and temperature differences.  For a single consistently defined
atmosphere, embedded fields with differing intrinsic processes must adjust
internally to find pressure balance with the overall average external
atmosphere.  In magnetically dominant regimes, the large-scale magnetic field
determines the density, temperature, and steady accelerated outflow everywhere
in the ambient atmosphere.

The temperature in static coupled equilibria follows the simple formula
\begin{equation}
kT_{\rm{e}}={\bar{m}g H_{\rm{Be}}\over2}={GM_{\rm{s}}\bar{m}H_{\rm{Be}}\over2R^2},
\label{e:corT}
\end{equation}
for mean particle mass $\bar{m}$, gravitational acceleration $g$, magnetic-field
scale height $H_{\rm{Be}}$, universal gravitational constant $G$, contained
object mass $M_{\rm{s}}$, and radial distance $R$ from the object center.  For a
given magnetic-field spherical-harmonic distribution, the scale height is
proportional to the object radius $H_{\rm{Be}}\sim R_{\rm{s}}$, giving coronal
temperatures above the object surface that go as the contained mass $M_{\rm{s}}$
over the object radius $R_{\rm{s}}$.  A near dipolar potential magnetic field
reproduces a correct solar coronal base temperature of 1.6E6K.  Supposing a
magnetic-field distribution similar to that of the sun, stars in the HR diagram
exhibit coronal temperatures that scale downward with increasing luminosity and
decreasing surface temperature from an average solar value of 1E6K down to 1E5K
right at the observed coronal dividing line, which is the cutoff temperature for
coronal spectral diagnostics.  Predicted coronal temperatures fall off beyond
the dividing line, consistent with observations of extended chromospheres in
cool giants and supergiants.  Similarly a temperature of 2E5K is predicted for
the average ambient background of the Milky Way, or a slightly lower temperature
of 1.6E5K at the solar galactic radial distance. The estimates are only a little
less than the observationally inferred $2-3$E5K widespread background gas
temperature.  If embedded current-carrying fields are present, neutron stars and
black holes are estimated to have exceedingly highly energized coronae, both in
the range of E11K = E4 keV.

Where magnetic fields are weakest over the globe, the coupled gas pressure is
reduced, which necessitates steady outflows, as are seen in solar coronal holes,
or near planetary and galactic poles.  Outflows add a compensating Bernoulli
pressure for equilibrium horizontal gravitational pressure stratification with
the surrounding static coupled coronal atmosphere.

As an allowed equilibrium, non-force-free flux tubes present at a boundary of an
otherwise large-scale coherent magnetic field can become embedded. Perturbed and
twisted non-force-free flux tubes expanding inhomogeneously with height in the
falling solar or stellar photospheric gas pressure may meet the lower coronal
magnetic-field boundary, like the chromospheric canopy, more-or-less intact to
become embedded flux tubes running parallel to the background near-potential
field in the corona above.  Sparsely distributed embedded non-force-free flux
tubes may represent only a small fraction of the total coronal volume and
magnetic-field energy.  Yet their tendency to relax to force-free all along
their lengths as an overall lowest energy state, may be opposed by magnetic
footpoint motions below the coronal base or by other processes originating
outside the coronal atmosphere.  Their footpoint motions or excitation
mechanisms may thereby supply the energy for the whole coronal atmosphere
against losses including those of intrinsic outflows.

\section*{Acknowledgements} 
For exploration and validation of derivations, Mathematica by Wolfram Research
Inc., Champaign IL, was utilized \citep{Wolfram1991}.  This research employed
NASA's Astrophysics Data System and the open access to historical collections
made available by many astronomical journals.

\appendix
\section*{Appendices}
\section{Small-Order Local Lorentz Force}
\label{a:Fi}

A full expansion of the Lorentz force gives clearer insight into the detailed
character of small deviations for a current-carrying field embedded in a varying
background magnetic field.  The remaining smaller-order terms in the expansion
of the Lorentz force from Eq.\ (\ref{e:F}) not contained in Eqs.\ (\ref{e:Fe0})
and (\ref{e:Fi0a}) or (\ref{e:Fi0b}) are rendered to ${\bf{F}}_{\rm{i1}}$ and
written out in this appendix.  The expressions ${\bf{F}}_{\rm{e}}$ and
${\bf{F}}_{\rm{i0}}$ in Eqs. (\ref{e:Fe0}) and (\ref{e:Fi0a}) are independent of
the coordinate system, but some of the remaining smaller-order terms in
${\bf{F}}_{\rm{i1}}$ are expanded using Cartesian vector identities.  To utilize
Cartesian vector identities requires that vectors be translated to Cartesian
coordinates beforehand.  Where results are restricted to less than 3 degrees of
freedom they are only applicable to Cartesian solutions.

The remaining terms from Eq.\ (\ref{e:F}) are written
\begin{equation}\begin{split} 
{\bf{F}}_{\rm{i1}}&={1\over4\pi}\Biggl(
\biggl\{{d\check{b}\over d\check{a}}\bigl({\bf\nabla}\check{a}^2{\bf\times}{\bf\Omega}\bigr)
\bigl({\bf\nabla}{\bf\times}{\bf\Omega}\bigr)\biggr\}_1
-\biggl\{\check{a}{\bf\nabla}^2\check{a}
\bigl({\bf\Omega}{\bf\times}({\bf\nabla}{\bf\times}{\bf\Omega})\bigr)\biggr\}_3\\
&\quad\quad-\biggl\{\bigl(2{\bf\nabla}\check{a}{\bf\cdot}{\bf\nabla}{\bf\Omega}+\check{a}{\bf\nabla}^2{\bf\Omega}\bigr){\bf\times}
\bigl({\bf\nabla}{\bf\times}(\check{a}{\bf\Omega})+\check{b}{\bf\Omega}\bigr)\biggr\}_{\top{3}{4}}\Biggr).
\label{e:Fi1a}
\end{split}\end{equation}
Applying the functional constraint $\check{b}=\check{b}[\check{a}]$ to the first
term in Eq.\ (\ref{e:F}) leaves just the term $\{\}_1$.  The second term in Eq.\
(\ref{e:F}) is fully expanded in ${\bf{F}}_{\rm{e}}$ and ${\bf{F}}_{\rm{i0}}$ in
Eqs. (\ref{e:Fe0}) and (\ref{e:Fi0a}) with the continuity of $\check{b}$ along
the lines of the flux surface from Eq.\ (\ref{e:divb1}).  The third term in Eq.\
(\ref{e:F}) leaves $\{\}_3$ and $\{\}_{\top{3}{4}}$ expanded using Cartesian
vector identities, and the fourth term in Eq.\ (\ref{e:F}) is contained in
$\{\}_{\top{3}{4}}$ also expanded using Cartesian vector identities.

In a potential background magnetic field or flux surface with
${\bf\nabla}{\bf\times}{\bf\Omega}=0$, the Lorentz force from Eq.\
(\ref{e:Fi1a}) simplifies to
\begin{equation}
{\bf{F}}_{\rm{i1}}=
-{1\over2\pi}\bigl({\bf\nabla}\check{a}{\bf\cdot}{\bf\nabla}{\bf\Omega}\bigr){\bf\times}
\bigl({\bf\nabla}\check{a}{\bf\times}{\bf\Omega}+\check{b}{\bf\Omega}\bigr),
\label{e:Fi1b}
\end{equation}
making use of the continuity of $\check{a}$ along lines of the flux surface from
Eq.\ (\ref{e:diva}) in $\{\}_3$, and the property of the potential field
${\bf\Omega}={\bf\nabla}\psi$ for a scalar potential function $\psi$ in
$\{\}_{\top{3}{4}}$, ${\bf\nabla}^2{\bf\Omega}={\bf\nabla}^2{\bf\nabla}\psi=
{\bf\nabla}{\bf\nabla}^2\psi={\bf\nabla}({\bf\nabla}{\bf\cdot}{\bf\nabla}\psi)=
{\bf\nabla}({\bf\nabla}{\bf\cdot}{\bf\Omega})=0$ for a Cartesian vector.

The leading factor in the single remaining term in ${\bf{F}}_{\rm{i1}}$ in Eq.\
(\ref{e:Fi1b}), ${\bf\nabla}\check{a}{\bf\cdot}{\bf\nabla}{\bf\Omega}$, can be
seen to vanish in current-sheet solutions.  The gradient of the continuity
relation for $\check{a}$ from Eq.\ (\ref{e:diva}),
${\bf\nabla}({\bf\nabla}\check{a}{\bf\cdot}{\bf\Omega})=0$, is expanded in
tensor notation for Cartesian geometry
\begin{equation}
0=\partial_{\imath}(\partial_{\jmath}\check{a}\partial_{\jmath}\psi)=
\partial_{\jmath}\partial_{\imath}\check{a}\partial_{\jmath}\psi
+\partial_{\jmath}\check{a}\partial_{\jmath}\partial_{\imath}\psi,
\nonumber
\end{equation}
using ${\bf\Omega}={\bf\nabla}\psi$, with vector element derivatives substituted
${\bf\nabla}\rightarrow\partial_{\imath}$, assuming implicit summations over a
repeated index in a term.  The expression represents the identity rewritten in
vector notation
\begin{equation}
({\bf\nabla}\check{a}{\bf\cdot}{\bf\nabla}){\bf\Omega}=
-({\bf\Omega}{\bf\cdot}{\bf\nabla}){\bf\nabla}\check{a}.
\label{e:curv0}
\end{equation}

The functional dependence for a most-significant-order isotropic Lorentz force
${\bf\nabla}^2\check{a}=({\bf\nabla}^2a)[\check{a}]$ can be seen to necessarily
imply ${\bf\nabla}\check{a}=({\bf\nabla}a)[\check{a}]$ in its three vector
elements for a 1D constituent function $\check{a}$ in Cartesian coordinates.
The functional dependence ${\bf\nabla}^2\check{a}=({\bf\nabla}^2a)[\check{a}]$
implies that the $2\times2$ Jacobian vanishes
\begin{equation}
{\partial({\bf\nabla}^2\check{a},\check{a})\over\partial(\xi_{\imath},\xi_{\jmath})}=
\partial_{\imath}\partial_k\partial_k\check{a}\partial_{\jmath}\check{a}
-\partial_{\imath}\check{a}\partial_{\jmath}\partial_k\partial_k\check{a}=0,
\nonumber
\end{equation}
for any pair of coordinates $\xi_{\imath}$ and $\xi_{\jmath}$.  Expanding the
derivative of a product, gives for the $2\times2$ Jacobian
\begin{equation}
\partial_k(\partial_{\imath}\partial_k\check{a}\partial_{\jmath}\check{a}
-\partial_{\imath}\check{a}\partial_{\jmath}\partial_k\check{a})=
\partial_{\imath}\partial_k\partial_k\check{a}\partial_{\jmath}\check{a}
-\partial_{\imath}\check{a}\partial_{\jmath}\partial_k\partial_k\check{a}=0,
\nonumber
\end{equation}
or in more conventional vector notation
\begin{equation}
\partial_k{\partial(\partial_k\check{a},\check{a})\over\partial(\xi_{\imath},\xi_{\jmath})}=
{\partial({\bf\nabla}^2\check{a},\check{a})\over\partial(\xi_{\imath},\xi_{\jmath})}=0,
\nonumber
\end{equation}
for any vector element $k$. Thus the Jacobian between ${\bf\nabla}\check{a}$ in
any of its vector elements and $\check{a}$ is a constant for any
$\imath\not=\jmath$.  However the function $\check{a}$ is localized by
assumption and must vanish with its derivatives away from the embedded field,
meaning that the constant must be zero
\begin{equation}
{\partial(\partial_k\check{a},\check{a})\over\partial(\xi_{\imath},\xi_{\jmath})}=0,
\nonumber
\end{equation}
in every vector element $k$ and for any pair of coordinates $\xi_{\imath}$ and
$\xi_{\jmath}$, which implies the functional dependence
${\bf\nabla}\check{a}=({\bf\nabla}\check{a})[\check{a}]$ in all three vector
elements.  It then follows from Eq.\ (\ref{e:curv0})
\begin{equation}
({\bf\nabla}\check{a}{\bf\cdot}{\bf\nabla}){\bf\Omega}=
-({\bf\Omega}{\bf\cdot}{\bf\nabla})(({\bf\nabla}\check{a})[\check{a}])
=-({\bf\Omega}{\bf\cdot}{\bf\nabla}\check{a}){d({\bf\nabla}\check{a})\over d\check{a}}=0,
\label{e:curv1}
\end{equation}
using the continuity of $\check{a}$ from Eq.\ (\ref{e:diva}).

For Cartesian current sheets embedded in a background potential magnetic field,
the leading factor in Eq.\ (\ref{e:Fi1b}) vanishes, and with it
${\bf{F}}_{\rm{i1}}$, using the functional constraint for Lorentz-force isotropy
from the most significant parts ${\bf{F}}_{\rm{e0}}$ and ${\bf{F}}_{\rm{i0}}$.
In this case the most significant Lorentz force ${\bf{F}}_{\rm{i0}}$ from Eq.\
(\ref{e:Fi0b}) is an exact isotropic solution.

In sheathed cylindrical flux-tube solutions ${\bf\nabla}\check{a}$ goes to zero
in the interior of the flux tube, away from the sheath, so the small-order
Lorentz force in embedded potential field solutions from Eq.\ (\ref{e:Fi1b})
also vanishes there.  In the sheath, a small-order anisotropic Lorentz force
$F_{\rm{i1}}$ of order ${\bf{B}}_{\rm{e}}^2/H_{\rm{Be}}$ is left compared to a
normal isotropic Lorentz force $F_{\rm{i0}}$ from Eq.\ (\ref{e:Fi0b}) of order
${\bf{B}}_{\rm{e}}^2/d_{\rm{sh}}$, giving Lorentz-force deviations in the thin
sheath of the small relative order $d_{\rm{sh}}/H_{\rm{Be}}$, for $d_{\rm{sh}}$
the thickness of the sheath.

\section{Solution Extensions}
\label{a:MHs1}

More general MHD equilibria with embedded fields separate into global and local
relations too, with the global relation resembling the original MHD equation
without the embedded fields, but with coupled gas and magnetic pressures.
Additional global terms in the MHD pressure Eq.\ (\ref{e:MHs1}) pass through
into the global pressure relation (\ref{e:MHs4b}) if they can be written as a
pressure $P$ multiplying a vector field, with the external pressure factor alone
varying in the vicinity of the embedded field, e.g.\ a flow viscosity term $\nu
(P/(H_{\rm{s}}g[R])){\bf\nabla}{\bf{u}}^2$ with constant viscosity $\nu$.

A local term that dies out away from the embedded field might add simply to the
local relation, if it is multiplied by a pressure factor $P$, like an additional
flow pressure
$(P/(H_{\rm{si}}g[r])){\bf{u}}_{\rm{i}}{\bf\cdot}{\bf\nabla}{\bf{u}}_{\rm{i}}$
or a variation in temperature or static pressure scale height in the
gravitational term $(P/H_{\rm{si}}){\bf\nabla}R$ local to the embedded field.
With the external pressure coupled to the flux surface
$P=\check{P}\exp[-\Pi]=\check{P}{\bf\Omega}^2$ (cf.\ Eqs.  (\ref{e:Pidefa}) and
(\ref{e:MHs3b})), the term becomes multiplied by ${\bf\Omega}^2$ and is absorbed
into the curly braces on the right side of Eq.\ (\ref{e:MHs2c}), adding to the
local relation Eq.\ (\ref{e:MHs3a}).

Mixed terms as due to ambient temperature differences or flows in the embedded
field, a Lorentz force due to a non-potential background magnetic field, or a
non-MHD anisotropic gas pressure, can be added too, but certain algebraic
manipulations may be required to separate them into a sum of separate global and
local terms.  For example, an ambient temperature difference is represented by a
gravitational term containing an offset in the static pressure scale height
$\Delta H_{\rm{s}}$ that is nonzero only local to the embedded field
\begin{equation}
{P\over H_{\rm{s}}-\Delta H_{\rm{s}}}{\bf\nabla}R=
{P\over H_{\rm{s}}}{\bf\nabla}R+{P\over H_{\rm{si}}}{\bf\nabla}R,
\quad{\rm{with}}\quad H_{\rm{si}}={H_{\rm{s}}-\Delta H_{\rm{s}}\over\Delta H_{\rm{s}}/H_{\rm{s}}}.
\nonumber
\end{equation}
The added gravitational term does indeed vanish away from the embedded field,
but $H_{\rm{si}}$ cannot be interpreted directly as an actual static pressure
scale height for an actual temperature in the vicinity of the embedded field.

With added local flow and gravitational terms in the MHD Eq.\ (\ref{e:MHs1}),
the MHD equation still separates, again applying the template Eq.\
(\ref{e:Pidefb}).  The global part of the MHD equation might still be equated to
${\bf\Omega}^2$ as in Eq.\ (\ref{e:MHs3b}), taking it to be the dominant global
factor in the Lorentz force for most of the embedded fields.  Small terms are
added to the local Lorentz force in the curly braces in Eq.\ (\ref{e:MHs2c}),
which introduce global variations in the local current-carrying field.  The
local pressure relation Eq.\ (\ref{e:MHs3a}) then becomes augmented as
\begin{equation}
{\bf\nabla}\check{P}=
-{1\over4\pi}\left(({\bf\nabla}^2\check{a})[\check{a}]+{1\over2}{d\check{b}^2\over d\check{a}}\right){\bf\nabla}\check{a}
+{\check{P}\over H_{\rm{si}}g[r]}{\bf{u}}_{\rm{i}}{\bf\cdot}{\bf\nabla}{\bf{u}}_{\rm{i}}
+{\check{P}\over H_{\rm{si}}}{\bf\nabla}R
+{{\bf{F}}_{\rm{i1}}\over{\bf\Omega}^2},
\label{e:MHs3ax}
\end{equation}
with local flows and ambient temperature differences, and including the
small-order $d_{\rm{cc}}/H_{\rm{Be}}$ Lorentz force ${\bf{F}}_{\rm{i1}}$ from
Eq.\ (\ref{e:Fi1a}) in Appendix \ref{a:Fi}.

The Lorentz force due to a global non-potential background magnetic field
${\bf{F}}_{\rm{e}}$ as defined in Eq.\ (\ref{e:Fe1}) adds consistently to the
separated equations too. Subtracting ${\bf{F}}_{\rm{e}}$ from both sides of Eq.\
(\ref{e:MHs1}) adds a global Lorentz force to the integrating factor in Eq.\
(\ref{e:MHs2a})
\begin{equation}
{\bf\nabla}\Pi={1\over H_{\rm{s}}g[R]}{D{\bf{u}}\over Dt}+{1\over H_{\rm{s}}}{\bf\nabla}R
-{\check{b}^2\over4\pi\check{P}{\bf\Omega}^2}
({\bf\nabla}{\bf\times}{\bf\Omega}){\bf\times}{\bf\Omega},
\label{e:MHs2ax}
\end{equation}
dividing by the pressure $P=\check{P}{\bf\Omega}^2$.  If the leading factor on
the global Lorentz force varies across the current-carrying field with
$\check{P}\not\sim\check{b}^2$ there, then the difference from a uniform
Lorentz-force term ${\bf{F}}_{\rm{e}}$ must be added to the local
current-carrying-field relation.

With the Lorentz force due to the non-potential magnetic field contributing to
the integrating factor for the global pressure in Eq.\ (\ref{e:MHs2ax}), the
MHD pressure balance follows from Eq.\ (\ref{e:MHs3b}), leading to an expanded
global relation Eq.\ (\ref{e:MHs4b}) having the expected form
\begin{equation}
{\bf\nabla}P_{\rm{e}}+{P_{\rm{e}}\over H_{\rm{s}}g[R]}{D{\bf{u}}\over Dt}+{P_{\rm{e}}\over H_{\rm{s}}}{\bf\nabla}R=
{1\over4\pi}({\bf\nabla}{\bf\times}{\bf{B}}_{\rm{e}}){\bf\times}{\bf{B}}_{\rm{e}},
\quad{\rm{with}}\quad
P_{\rm{e}}=\beta_{\rm{c}}{{\bf{B}}_{\rm{e}}^2\over 8\pi},
\label{e:MHs4x}
\end{equation}
using the asymptotic relations ${\bf{B}}_{\rm{e}}=\check{b}_e{\bf\Omega}$ and
$P_{\rm{e}}=\check{P}{\bf\Omega}^2$.

Though the cosmic-ray pressure may be quite strong, it is commonly taken to be
independently coupled to the background magnetic field (\citealt{ParkerEN1966b},
\citealt{Ames1973}, or see discussion \citealt{Shu1974}).  This leaves the
gas-magnetic pressure coupling and equilibrium relations unaltered in form as
further elaborated in other work to be presented.


\end{document}